\documentclass[preprintnumbers,amsmath,12pt,amssymb,floatfix,superscriptaddress,nofootinbib]{article}

\def\mytitle#1{\setcounter{equation}{0}
\setcounter{footnote}{0}
\begin{center}\Large\textbf{#1}\end{center}
\vspace{0.25cm}}
\def\myname#1{\begin{center}{\large #1}\end{center}\vspace{-0.13cm}}
\def\myplace#1#2{\small\begin{center}\textit{#1}\\
\texttt{#2}\end{center}}

\def\myclassification#1{\small\noindent
Keywords :
       #1\vspace{0.5cm}}
\usepackage{graphicx}% Include figure files
\usepackage{amssymb}
\usepackage{amsmath}
\usepackage{graphicx}
\usepackage{epstopdf}
\usepackage{hyperref}
\usepackage{color}
\usepackage{xcolor}
\usepackage{subfig}
\usepackage{multirow}
\usepackage{array}
\begin{document}

\mytitle{Stability Analysis of Four $f(Q)$ Gravity Models : A Cosmological Review in the Background of Bianchi-I Anisotropy}

\myname{$Subhajit~Pal^*$\footnote{subhajitpal968@gmail.com $~~;~~\text{Orchid}~:~0009-0002-2431-1375$}, $Atanu~Mukherjee^{*}$\footnote{atanu2002mukherjee@gmail.com$~~;~~\text{Orchid}~:~0009-0003-3230-0059$}, $Ritabrata~
Biswas^{**}$\footnote{biswas.ritabrata@gmail.com$~~;~~\text{Orchid}~:~0000-0003-3086-892X$} and $Farook~Rahaman^*$ \footnote{farookrahaman@gmail.com$~~;~~\text{Orchid}~:~0000-0003-0594-4783$}} 

\vspace{0.5cm}
\myplace{*Department of Mathematics, Jadavpur University, Kolkata-32, India\\ **Department of Mathematics, The University of Burdwan, Burdwan-713104, India} {}
%%%%%%%%%%%%%%%%%%%%%%%%%%%%%%%%%%%%%%%%%%%%%%%%%%%%%%%%%%%%%%%%%%%%%%%%%%%%%%%%%%%%
\begin{abstract}
With the non-metricity scalar $Q$ as the functional argument, several $f(Q)$ gravity models are found to be proposed which are perfectly able to mimic the late-time accelerated expansion as pointed out by the type Ia supernovae observations. Temperature fluctuation differences for two celestial hemispheres, Hubble tension, voids, dipole modulation, anisotropic inflation, etc. motivates us to think beyond the $\Lambda$CDM model and the cosmological principle. Bianchi-I model portrays an anisotropic universe imposing shear. $f(Q)$ model also enables us to produce early inflation to late de Sitter universe without the requirement of $\Lambda$CDM. Ambiguities regarding fine-tuning or coincidences can be avoided alongwith. So, this article finds different stationary points of cosmic evolution with $f(Q)$ models habilitating in Bianchi-I anisotropic universe. Depending on models' nature, fixed points with different categories are found. Perturbations are followed wherever are applicable. While pursuing cosmological implications towards these fixed points, some are found to be formed only for the consideration of $f(Q)$ gravity and Bianchi-I both. Besides different prediction towards early inflation to late-time expansion which are available in existing literature of dynamical system studies, occurances of ultra slow roll inflation is predicted. For particular $f(Q)$ model, shear is predicted to decay leaving behind a constant valued residue. This models a universe that gradually turns more homogeneous. In some other models, depending on initial conditions, a final isotropic leftover is marked as the future fate of anisotropic world. More than one stable points are marked for special cases and are cosmologically interpreted. 
\end{abstract}
%%%%%%%%%%%%%%%%%%%%%%%%%%%%%%%%%%%%%%%%%%%%%%%%%%%%%%%%%%%%%%%%%%%%%%%%%%%%%%%%%%
\myclassification{Anisotropic universe, Dynamical system, Center manifold theory}
%%%%%%%%%%%%%%%%%%%%%%%%%%%%%%%%%%%%%%%%%%%%%%%%%%%%%%%%%%%%%%%%%%%%%%%%%%%%%%%%%
\section{Introduction}
%%%%%%%%%%%%%%%%%%%%%%%%%%%%%%%%%%%%%%%%%%%%%%%%%%%%%%%%%%%%%%%%%%%%%
As explained by the cosmological principle, our universe is postulated to be both homogeneous and isotropic on a large scale. This generates a framework described by the Friedmann–Lemaître–Robertson–Walker (FLRW) geometry. In recent years, observations of the universe's expansion rate and structure at early times (e.g., from the Cosmic Microwave Background (CMB)) and late-times (e.g., from supernovae, baryon acoustic oscillations (BAO) and galaxy surveys) have shown inconsistencies or ``tensions". One conspicuous example is the Hubble tension. Early universe measurements from the Planck satellite, assuming standard cosmological model that describes the universe, incorporating a cosmological constant (represented by the Greek letter lambda, $\Lambda$) associated with dark energy(DE), and cold dark matter (CDM) $-$ 
$\Lambda$CDM, speculate the present day value of Hubble parameter, $H_0\approx 67.4\pm 0.5 ~km~s^{-1}/Mpc$ \cite{aghanim2020planck}. On the contrary, from the  late universe measurements, predicted from type Ia supernovae observation, the project Supernova, $H_{0}$, for the Equation of State (SH0ES collaboration), the value of $H_0$ is found to be $\approx 73.2\pm 1.3~ km ~s^{-1}/Mpc$ \cite{Riess2022H0}. The discrepancy between these two measurements is the most prominent tension in the context of the Hubble tension. These tensions suggest that the standard $\Lambda$CDM model which assumes general relativity(GR) and the cosmological principle, might be incomplete or there must be a missing key of physics.

Alternative scenario thus required involves either modified gravity theories or relaxation on the cosmological principle considering universes that are not perfectly homogeneous or isotropic, especially in the early stages. This leads to different models like Bianchi cosmologies (anisotropic but homogeneous) and Lemaître–Tolman–Bondi (LTB) models (inhomogeneous).

Observationally, anisotropies and inhomogeneities are supported from the cosmic microwave background (CMB) radiation which is the relic radiation from the surface of last scattering about $380,000$ years after the Big Bang when photons were decoupled from matter\cite{Planck2018overview}. It is followed that the quadrupole $(l=2)$ moment of the CMB power spectrum is lower than that predicted by $\Lambda$CDM. Its statistical significance alone is not strong $(\sim 2-2.5\sigma)$ but it is a part of a set of correlated anomalies\cite{aghanim2020planck}. Wilkinson Microwave Anisotropy Probe (WMAP) followed that the amplitude of temperature fluctuations appears stronger in one hemisphere of the sky compared to the other\cite{eriksen2004asymmetries}. A large, unusually cold region, centered near galactic coordinates $(l,b)\approx (209^{o},-57^{o})$, was detected in WMAP and confirmed by Planck \cite{kovacs2014supervoid}. These dipole modulation, anisotropic inflation, void etc point towards anisotropies.

%%%%%%%%%%%%%%%%%%%%%%%%%%%%%%%%%%%%%%%%%%%%%%%%%%%%%%%%%%%%%%%%%%%%%
%\section{Basic formalism of $f(Q)$ gravity}
%%%%%%%%%%%%%%%%%%%%%%%%%%%%%%%%%%%%%%%%%%%%%%%%%%%%%%%%%%%%%%%%%%%%%
Now, we will consider the metric-affine geometry where the metric $g_{\mu \nu}$ and the affine connection $\Gamma_{\mu \nu}^\lambda$ are independent. If we impose both the curvature and the torsion to vanish,  i.e., $R_{\mu\nu\rho}^\lambda=~T_{\mu \nu}^\lambda=0$, the non-metricity is only allowed not to vanish identically. The non-metricity tensor takes the form :
\begin{equation}
    Q_{\lambda \mu \nu} = \nabla_{\lambda} g_{\mu \nu} \neq 0~~~.
\end{equation}
We define two traces :
\begin{equation}
    Q_\lambda = Q_{\lambda \mu}^\mu~~~~~\text{and}~~~~~~\bar{Q}_\lambda =Q_{\mu \lambda}^\mu~~~~.
\end{equation}
The superpotential tensor $P_{\mu \nu}^\lambda$ is given by,
\begin{equation}
    P_{\mu \nu}^\lambda=\frac{1}{4}\left({-Q_{\mu \nu}^\lambda+ 2 Q_{(\mu~\nu)}^\lambda- Q^\lambda g_{\mu \nu}-\bar{Q}^\lambda g_{\mu \nu}}\right)~~~.
\end{equation}
This has a similar role to the Einstein's tensor in GR.

Using the superpotential tensor, we can form a scalar in connections with the non-metricity tensor given by,
\begin{equation}
    Q=- g^{\mu \nu} \left({P_{\mu \nu}^\lambda Q_\lambda^{\mu \nu}}\right)~~~~.
\end{equation}

This non-metricity scalar $Q$ plays almost a similar role as the Ricci scalar $R$ plays in GR.

The action of $f(Q)$ gravity is given by,
\begin{equation}\label{5}
    \mathcal{S} = \int {\sqrt{-g}\left[{\frac{1}{2} f(Q) + \mathcal{L}_m}\right] d^4 x}~~~~,
\end{equation}
where $f(Q)$ is the arbitrary function of the non-metricity scalar $Q$ and $\mathcal{L}_m $ is the matter Lagrangian.

Varying the action with respect to the metric $g_{\mu \nu}$, equation (\ref{5}) gives, 
\begin{equation}
    \frac{2}{\sqrt{-g}} \nabla_\lambda \left({\sqrt{-g} f_Q P_{\mu \nu}^\lambda}\right)+\frac{1}{2} g_{\mu\nu} f(Q) + f_Q \left({P_{\mu \lambda \rho} Q_{\nu}^{\lambda \rho} - 2 Q_{\lambda \rho \mu} P_\nu^{\lambda \rho}}\right)= T_{\mu \nu}~~~~,
\end{equation}
where $f_{Q} \equiv \dfrac{d}{dQ} f(Q)$ and $T_{\mu \nu}$ is the energy-momentum tensor.

%%%%%%%%%%%%%%%%%%%%%%%%%%%%%%%%%%%%%%%%%%%%%%%%%%%%%%%%%%%%%%%%%%%%%
%\section{Bianch-I cosmology.}
The Bianchi-I metric is the simplest anisotropic generalization of the flat FLRW metric. This describes a spatially homogeneous but anisotropic universe where expansion can vary in different spatial directions. The metric in comoving coordinates $(t,~x,~y,~z)$ is given as, 
\begin{equation}\label{FLRW spacetime metric}
    ds^2 = - dt^2 +\sum_{i=1}^3 a_i^2 (t) (dx^i)^2~~~~,
\end{equation}
where $a_1^2(t)$, $a_2^2 (t)$ and $a_3^2 (t)$ respectively are the directional scale factors describing how distances expand respectively along the $x,~y$ and $z$ axes. There is no spatial dependence in the metric function. Only a time dependence is followed. This means every point in space is geometrically identical at a given time. The scenario turns homogeneous as a result. However, the inequality of $a_1(t)$, $a_2(t)$ and $a_3(t)$ indicates different rates of expansion or contraction in different directions.

For this current situation, in anisotropic Bianchi-I space-time metric (\ref{FLRW spacetime metric}), the directional Hubble parameters are given by, $H_i =\frac{\dot{a_i}}{a_i}~,~\cdot \equiv \frac{d}{dt}$. Hubble parameter can be defined by the arithmetic mean of these three directions' Hubble parameters, given by,
\begin{equation}
    H_{AM}(t) = \frac{1}{3} \left[{H_1 +H_2 +H_3}\right]~~.
\end{equation}
We also denote the scale factor $a_{GM}(t)$ as the geometric mean of the directional scale factors, given by,
\begin{equation}
    a_{GM}(t) = \left[{a_1(t) a_2(t) a_3(t)}\right]^{\frac{1}{3}}~~.
\end{equation}
Now, we can use the parametrisation, 
\begin{equation}
    a_i(t) = a_{GM}(t) e^{\beta_i (t)}~~.
\end{equation}
Then calculating we get,
\begin{equation}
    H_i = H+ \beta_i~,~~~~\beta_1 + \beta_2 +\beta_3 =0~~.
\end{equation}
The space-time metric admits three isometries associated with the vector fields $\partial_x,~\partial_y$ and $\partial_z$ respectively.

The Bianchi-I geometry, given in Cartesian coordinates, possesses the nonzero components of the three Christoffel connections as,
\begin{align*}
    \Gamma_1~&:~\Gamma_{tt}^t =\gamma(t)~,\\
    \Gamma_2~&:~\Gamma_{tt}^t = \gamma(t) +\frac{\dot{\gamma}(t)}{\gamma(t)}~,~~\Gamma_{tx}^x = \Gamma_{ty}^y \Gamma_{yz}^z= \gamma(t)~~~\text{and}\\
    \Gamma_3~&:~\Gamma_{tt}^t = -\frac{\dot{\gamma}(t)}{\gamma(t)}~,~~\Gamma_{tx}^t = \Gamma_{ty}^t \Gamma_{yz}^t = -\gamma(t)~~,
\end{align*}
where the function $\gamma$ introduces dynamic degrees of freedom in the field equation.

The non-metricity scalar $Q$, derived from the connection of $\Gamma_1$ is obtained as, \cite{Paliathanasis2024, murtaza2025generic}
\begin{equation}\label{expression of Q in gamma_1 connection}
    Q = 2 \sum H_i H_j = 6 H^2 - \sigma^2~~,
\end{equation}
\begin{equation}
    \text{where}~~~\sigma^2 \equiv \frac{1}{2}\,\sigma_{\mu\nu}\sigma^{\mu\nu}
= \frac{1}{2}\sum_{i=1}^{3}\left(H_i - H\right)^{2}= \sum \dot{\beta}^2 ~~\text{with}~~H_i \equiv \frac{\dot a_i}{a_i}, \quad H \equiv \frac{1}{3} \sum_{i=1}^{3} H_i ..
\end{equation}

Note that $\sigma=0 $ implies $\sum\dot {\beta}^2=0$ which means the universe is an isotropic FLRW.

In symmetric teleparallel gravity, the sign of the non-metricity scalar $Q$ depends on the background geometry and the relative contributions of expansion and shear. 

Consequently, while $Q$ is typically positive in expansion-dominated regimes, it may become negative in strongly anisotropic phases where the shear contribution dominates ($\sigma^2 > 6H^2$). Since several of the $f(Q)$ models considered in this work involve powers, square roots, or logarithms of $Q$, the analysis is restricted to the region of phase space where $Q>0$, corresponding to physically relevant solutions with moderate anisotropy.

This assumption ensures that all $f(Q)$ functions and their derivatives are well-defined. The fixed points and phase-space trajectories discussed below are therefore implicitly understood to lie within the $Q>0$ sector of the theory.
Since the value of $H^2$ and shear $\sigma^2$ always positive, so the sign of $H$ and the shear does not depend on the nature of $Q$.

The anisotropic fluid is defined as, 
\begin{equation}
    T_\mu^\nu = diag (-\rho,~P_1,~P_2,~P_3) = diag(-\rho,~ \omega_1 \rho,~ \omega_2 \rho,~ \omega_3 \rho)~~,
\end{equation}
where $\rho$ denotes the energy density of the fluid, $P_1,~P_2~\text{and}~P_3$ are the pressure along the three orthogonal directions. The directional equation of state (EoS) parameters are $\omega_1,~\omega_2~\text{and}~\omega_3$. Then the average EoS parameter $\omega$ is given by,
\begin{equation}
    \omega= \frac{1}{3} \left({\omega_1+\omega_2+\omega_3}\right)~~,
\end{equation}
and the deviations $\mu_i$ can be calculated from the average EoS as,
\begin{equation}
    \mu_i = \omega_i-\omega~,~~~\text{or},~\omega_i = \omega+ \mu_i~,~~~\text{for}~i=1,2,3~.
\end{equation}
However, for the isotropic fluid, $P_1=P_2=P_3=P= \omega \rho$.

Now, the modified Friedmann and the Raychaudhuri equations for our case can be expressed as(for simplification, writting $H_{AM}$ as $H$ and $a_{GM}$ as $a$),
\begin{equation} \label{modified Friedmann equation}
    3H^2 -\sigma^2 = \frac{\kappa}{f_Q} \left[{\rho - \frac{3 H^2 f_Q}{\kappa}-\frac{f}{2 \kappa}}\right]~~\text{and}
\end{equation}
\begin{equation}\label{Modified Raychaudhuri equation}
    -(2 \dot{H} +3 H^2) -\frac{\sigma^2}{2} = \frac{\kappa}{f_Q} \left[{\omega_\rho +\frac{f}{2 \kappa}+\frac{2H \dot{f}_Q}{\kappa}-\frac{Qf_Q}{2 \kappa}}\right]~~.
\end{equation}
The equations \eqref{modified Friedmann equation} and \eqref{Modified Raychaudhuri equation} can be simplified, taking into account the expression for $Q$ from \eqref{expression of Q in gamma_1 connection}, as
\begin{align} \label{reduced friedmann equation}
    3 H^2 -\frac{\sigma^2}{2} &= \frac{\kappa}{f_Q} \left[{\rho -\frac{f}{2 \kappa}}\right]~~and\\ 
    -(2 \dot{H} +3 H^2) -\frac{\sigma^2}{2} &= \frac{\kappa}{f_Q} \left[{\omega_\rho +\frac{f}{2 \kappa}+\frac{2H \dot{f}_Q}{\kappa}}\right]~~,~~~\text{where}\\ \label{reduced Raychaudhuri equation}
    \dot{\sigma} &=-\sigma \left({\frac{\dot{f}_Q}{f_Q} + 3H}\right)~~.
\end{align}
The last equation is known as an anisotropic evolution equation. Integrating we have,
\begin{equation}
    \sigma \sim \frac{1}{a^3 f_Q}~~~.
\end{equation}

Locally Rotationally Symmetric (LRS) Bianchi type-I universe within the framework of a novel modified $f(Q)$ gravity theory is studied in the article \cite{Rathore2024}. In this theory, gravitational interactions are chosen to be governed by the non-metricity tensor $Q$, under the assumption of vanishing curvature and torsion. Fixed points are found to signify quintessence-like DE cosmological scenario as well as the phantom DE cosmological model. Authors of the article \cite{Paliathanasis2024} focused on solutions defined in the coincident gauge where the authors identify Kasner-like solutions that is built upon the original Kasner relations. Interestingly, the characteristics of the indices for these exact solutions remain consistent which suggests that they exhibit similar behavior to that of the Kasner universe. On the other hand, in the same work, other two sets of field equations defined by connections in the non-coincident gauge along with  self-similar solutions are found that do not adhere to the Kasner-like constraint equations. This observation implies that the chaotic behavior typically seen in the Mixmaster universe, especially near singularities, which may not be present in the context of symmetric teleparallel $ f(Q) $-gravity. 

General forms of homogeneous and isotropic symmetric teleparallel geometries are identified in the article \cite{Hohmann2021}. One solution with spatial curvature (where $ k \neq 0$) is followed to meet the symmetry conditions. Three distinct branches of spatially flat geometries (where $ k = 0 $) are noted, each of which can be derived as a specific limit from the common spatially curved case as $ k $ approaches zero. Each of these branches is characterized by two scalar functions of time, viz., the cosmological scale factor $ a$ in the metric and another scalar $ K $ that describes the symmetric teleparallel connection. A third function, the lapse $ N $, can be adjusted through a choice of time coordinate. For each branch,  the coincident gauge  was found where the coefficients of the symmetric teleparallel connection vanish.
Only one of these branches retains the Friedmann-Lemaître-Robertson-Walker (FLRW) form in the coincident gauge. 

Cosmological dynamical systems that emerge from various modified theories of gravity are explored model independently in the article \cite{bohmer2022cosmological} by concentrating on theories characterized by second-order field equations such as those in $ f(G) $, $ f(T) $, and $ f(Q) $ gravity theories.  For instance, in $ f(R) $ gravity, the Ricci scalar contains terms that are proportional to both $ H $ and $ \dot{H} $, which means that the variables chosen are not sufficient to fully close the system. The system will always be closed once a function $ f $ is specified, because any function of $ G $, $ T $ or $ Q $ can be expressed as a function of $ H $. This flexibility also extends to non-minimal couplings of matter, as long as the equations remain of second order.

The article \cite{shabani2023phase} analyzes $f(Q)$ gravity in spatially flat FLRW space-time using dynamical systems, revealing stable de Sitter solutions and unstable matter-dominated phases, highlighting the role of non-vanishing affine connections. Authors of the articles \cite{paliathanasis2023dynamical,shabani2024cosmology, esposito2022bianchi} have investigated a non-flat FLRW universe, finding curvature driven inflationary points and solutions alleviating the coincidence problem. The research of \cite{esposito2022bianchi} explores Bianchi-I cosmologies in $f(Q)$ gravity using the $1+3$ covariant formalism and dynamical system analysis. The article \cite{Rathore2024} analyzes the Locally Rotational Symmetric (LRS) Bianchi-I model in $f(Q)$ gravity. Also it links the saddle and stable critical points to quintessence and phantom DE scenarios. Also, finding the heteroclinic connections between radiation, matter and dark energy dominated phases are explored in \cite{sarmah2024dynamical}. Effective field theory on dark energy also effects on modified gravity, are reviewed in the article \cite{frusciante2020effective}.

Recently, a generic dynamical-system formulation for Bianchi-I cosmology with an isotropic fluid in $f(Q)$ gravity has been developed by Murtaza and Chakraborty~\cite{murtaza2025generic}. In particular, for the $\Gamma_{1}$ connection class, their autonomous system and constraint equations coincide with those obtained independently in the present work. We emphasise that our analysis is fully consistent with this framework, which we adopt here as a basis for a detailed stability and physical viability study of specific $f(Q)$ models.

The modified gravity can contribute to explaining anomalies in Cosmic Microwave Background(CMB) like the quadrupole suppression or axis of evil. Study of $f(Q)$ gravity offers new candidates for dark energy. This theory may avoid fine-tuning problems of $\Lambda$CDM. Bianchi-I background enables detailed perturbation analysis in anisotropic settings which inform $N$-body simulations in non-standard cosmologies. This study of $f(Q)$ can investigate how non-metricity drives inflation in anisotropic setups. It can fit anisotropic models with SNe Ia, BAO, CMB.

Dynamical system studies for $f(Q)$ gravity in the background of the Bianchi-I type anisotropic universe is important for theoretical, observational and cosmological reasons. This model is expected to probe the early universe anisotropies good. CMB observations suggest the possibility of small anisotropies or preferred directions. Investigating $f(Q)$ gravity in this context helps us to test whether geometric generalizations like symmetric teleparallelism can naturally resolve or damp anisotropies. A viable modified gravity theory should reproduce isotropy at late-times and allow anisotropy at early times. Dynamical systems help to test this evolution. GR
tends to isotropize the universe at late-times (Wald's theorem), but it is not guaranteed in modified theories. $f(Q)$ gravity modifies the connection based description of gravity. The dynamical system analysis mass reveal whether $f(Q)$ models allow natural isotropization. A dynamical system approach may help classifying models according to late-time cosmic acceleration, matter dominated era, stability and robustness analysis etc. \cite{Harko2018, Frusciante2021, CapozzielloDialektopoulos2021}

In the next section, we will construct general dynamical system equations for $f(Q)$ gravity in Bianchi-I cosmology. In section 3, we will analyze dynamical systems for different $f(Q)$ models. Phase portraits will be thoroughly checked and different equilibrium points will be physically interpreted. Finally, in section 4, a brief discussion of the work will be incorporated.
%%%%%%%%%%%%%%%%%%%%%%%%%%%%%%%%%%%%%%%%%%%%%%%%%%%%
\section{Dynamical Systems for Some $f(Q)$ Candidates in Bianchi-I Cosmology}

Define the dimensionless variables as, \cite{murtaza2025generic}
\begin{equation}\label{dynamical variables}
    x= \frac{\kappa \rho}{6 f_Q H^2}~,~~~~y = -\frac{f}{12 f_Q H^2}~~~\text{and}~~~~z = \frac{\sigma^2}{6H^2}~~.
\end{equation}

These variables admit a clear physical interpretation: $x$ measures the matter contribution to the cosmic expansion, $y$ encodes the effective geometric contribution arising from the $f(Q)$ modification (playing the role of an effective dark-energy component), and $z$ quantifies the degree of anisotropy in the form of shear. The isotropic FLRW limit corresponds to $z=0$, while anisotropic solutions are characterised
by $z>0$.

Here, $x$ is effective matter fraction. This parameter measures the robustness of cosmic expansion which is drawn by ordinary matter. Presence of $f_Q$ enforces the $\rho$ to gravitate differently. If $x$ is large, we expect decelerated expansion. Whereas $x\rightarrow 1$ signifies matter dominated era.

The second parameter $y$ signifies the geometric or DE sector. This encodes purely geometric contribution from the modified gravity Lagrangian. This stays constant when we choose cosmological constant type universe. If evolving DE is opted, $y$ turns dynamical. 

$z$ quantifies the degree of anisotropy in the form of shear. The isotropic FLRW limit corresponds to $z=0$, while anisotropic solutions are characterised
by $z>0$. Thermodynamically, $x$ is linked to energy conservation. $y$ is linked to entropy and horizon thermodynamics(especially in $f(Q)$) and $z$ is linked to dissipation and instability-low anisotropy distorts expansion.

Constrained by the relation \eqref{reduced friedmann equation},
\begin{equation}\label{dynamical variable constraint}
    x+y+z =1~~~.
\end{equation}

For the $\Gamma_{1}$ connection class, the dimensionless variables and the resulting autonomous dynamical system obtained in the present work are algebraically identical (up to notation) to those presented in Ref.~\cite{murtaza2025generic}. In particular, our variables $(x, y, z)$ correspond directly to their $(x_{1}, x_{2}, x_{3})$, and the associated constraint and evolution equations map term-by-term onto each other. We briefly summarise the formulation here for completeness and to establish our notation before proceeding with the model-specific analysis.

Now, the dynamical system equations can be constructed by taking the derivatives of the dimensionless variables $x,~y$ and $z$ with respect to the variable $N= ln~a$, given as, 
\begin{equation}
    x^\prime = -3x (1+\omega)+6x \left\{{\frac{(y-1-x \omega)(1+\Gamma -\Gamma z -2z)}{(\Gamma+2)(z-1)}}\right\}-\frac{6xz}{(\Gamma+2)(z-1)}~~,
\end{equation}
\begin{equation}
    y^\prime = 3 \Gamma (\frac{z+y-1-x\omega}{\Gamma+2}) + 6y \left[{\frac{(y-1-x \omega)(1+\Gamma -\Gamma z -2z)}{(\Gamma+2)(z-1)}}\right]-\frac{6yz}{(\Gamma+2)(z-1)}~~\text{and}
\end{equation}
\begin{equation}
    z^\prime = 6 z (x\omega-y)~~~,
\end{equation}
where ``$(~^\prime)$" denotes the derivative w.r.to $N=ln~a$. Then we have the auxiliary quantity $\Gamma=\Gamma(Q)$ as, 
\begin{equation}\label{Gamma equation}
    \Gamma = \frac{f_Q}{Qf_{QQ}}~~~.
\end{equation}
Now, by using \eqref{reduced Raychaudhuri equation}, we have,
\begin{equation}\label{H equation}
    \frac{\dot{H}}{H^2} = 3 \left\{{\frac{(2z+z\Gamma -\Gamma)(y-1-x \omega) + 2z}{(\Gamma+2)(z-1)}}\right\}~~~.
\end{equation}
Eliminating the variable $y$, the above dynamical system will be reduced to,
\begin{align}
    x^\prime &= - 3 x(1+\omega) + 6 x (x+z+x \omega)+ 6 x \frac{x+x \omega}{(\Gamma +2)(z-1)}~~~~\text{and} \\
    z^\prime &= 6 z x (1+\omega) +6 z^2 -6 z~~~,
\end{align}
while the equation \eqref{H equation} is reduced to,
\begin{equation}\label{Reduced H equation}
    \frac{\dot{H}}{H^2} = 3 \left\{{\frac{(2z+z\Gamma -\Gamma)(-z-x-x \omega) + 2z}{(\Gamma+2)(z-1)}}\right\}~~~.
\end{equation}

The equation \eqref{Reduced H equation} helps us to determine the cosmic evaluation for a particular fixed point.

To support the qualitative cosmological interpretations discussed above, we explicitly evaluate the effective equation of state parameter 
\begin{equation}
\omega_{\mathrm{eff}} \equiv -1 - \frac{2}{3}\frac{\dot H}{H^{2}},
\end{equation}
at the relevant fixed points for each of the four $f(Q)$ models considered in this work.

For a $f(Q)$ gravity model, we need to express $Q$ and hence $\Gamma(Q)$ in terms of the dynamical variables $x$ and $z$. Now from the equations \eqref{expression of Q in gamma_1 connection} and \eqref{dynamical variables}, we can construct the value of $Q$ as,
\begin{equation}
    \frac{Qf_Q}{f} =\frac{(1-z)}{2y}~~~.
\end{equation}
By using the expression of the equation \eqref{dynamical variable constraint}, one can derive,
\begin{equation}\label{expression of Q}
    \frac{Qf_Q}{f} =\frac{(1-z)}{2(1-x-z)}~~~.
\end{equation}

The expressions for $Q f_{Q}/f$ and their representation in terms of the dynamical variables follow directly from the generic formulation of Ref.~\cite{murtaza2025generic}, and are reproduced here for convenience before applying them to the specific $f(Q)$ models studied in this work.

Thus, we can say, from the above equations, that $Q=Q(x,~z)$. By using the expression of $\Gamma(Q)$, one can obtain an autonomous dynamical system for the connection $\Gamma_1$ for the Bianchi-I cosmology with an isotropic fluid in the modified gravity or $f(Q)$  gravity.

While the generic dynamical system structure for Bianchi-I cosmology in $f(Q)$ gravity is shared with Ref.~\cite{murtaza2025generic}, the novelty of the present analysis lies in the following aspects:
\begin{itemize}
    \item a systematic stability analysis of four specific $f(Q)$ models that have not been jointly analysed in the Bianchi-I context previously;
    \item the identification of multiple de~Sitter branches, non-hyperbolic critical points, and their cosmological interpretation in an anisotropic background;
    \item a detailed perturbative viability discussion, including ghost conditions, tensor propagation speed $c_T$, and sound-speed constraints, which goes beyond background dynamical analysis;
    \item a qualitative discussion of shear decay, isotropisation, and bifurcation structure that is specific to the models considered in this work.
\end{itemize}

We emphasise that the present work is intended as a systematic comparative analysis of specific $f(Q)$ models in an anisotropic Bianchi-I background, rather than the introduction of fundamentally new dynamical systems techniques. The dynamical framework employed here builds on existing formulations, and our focus is on extracting physical insight from stability properties, shear evolution, and viable cosmological trajectories within this setting. We stress that the perturbative viability conditions considered here are based on extensions of the isotropic FLRW results, and are employed as necessary (though not sufficient) conditions for physical consistency in the anisotropic background. A fully rigorous perturbation analysis in Bianchi-I spacetime is beyond the scope of the present work and is left for future investigation.

In the next section, phase portrait analysis of individual $f(Q)$ models chosen will be performed.

%%%%%%%%%%%%%%%%%%%%%%%%%%%%%%%%%%%%%%%%%%%%%%%%%%%%%
\section{Stability Analysis for Different $f(Q)$ Gravity Candidates}

In this section, we calculate the stability of the different $f(Q)$ gravity models for the Bianchi-I in the connection $\Gamma_1$. To perform this analysis in the presence of the matter component and will plot the phase portraits for every model.
%%%%%%%%%%%%%%%%%%%%%%%%%%%%%%%%%%%%%%%%%%%%%
\subsection{Model I : $f(Q) = m Q^n$}
%%%%%%%%%%%%%%%%%%%%%%%%%%%%%%%%%%%%%%%%%%%%%%%
The model $f(Q) = mQ^n$ is known as a power law model with the parameters $m$ which sets the strength of the gravitational computing and $n$ which determines how non-linearly the gravity theory depends on the scalar $Q$. At $n=1$, this model reduces to $f(Q)= mQ$ which is equivalent to GR if $m=1$. For specific powers $n>1$, this model can mimic inflationary potentials without incorporating scalar fields like the inflation in standard models \cite{mandal2023accelerated}. With $n=2$, Starobinsky inflation in $f(R)=R+R^2$ is mimiced which is found by Planck CMB data. The second order field equations make it easier to avoid instabilities common in higher derivative inflationary model. This model also modifies the tensor to scalar ratio $r$, affecting gravitational wave signals from inflation \cite{Khyllep2021}. For $0<n<1$, the modifications to gravity weakens gravitational attraction at large scales, mimicking a repulsive force. This is indicative towards the late-time cosmic acceleration. In general, a gravity model affects the generation and propagation, amplitude decay, polarization of gravitational waves(GWs). After the GW170817 event \cite{2017ApJabott} confirmed GWs travel at light's speed, $f(Q)=mQ^n$ gravity got a huge support as this too predicts GW's speed to be equal to that of light in vacuum. $f(Q)=mQ^n$ modifies the Poisson equation and growth equation for matter density perturbation given by $\ddot{\delta}+2h\dot{\delta}-4\pi G_{eff} \rho\delta=0$, where $\delta\equiv \frac{\delta\rho}{\rho}$ is the matter density contrast and $G_{eff}$ is the effective gravitational constant which can differ from Newton's $G_N$ in modified gravity theories like $f(Q)$ gravity. This affects the growth rate $f=\frac{d\ln \delta}{d \ln a}$ and the matter power spectrum\cite{Khyllep2021}. This predicts  towards different growth and structure formation histories compared to $\Lambda$CDM which is testable via redshift-space distortion, gravity clustering weak lensing surveys etc\cite{Khyllep2021}.

Now, the $Q$ derivatives of the model-I can be obtained as,
\begin{equation}\label{Model I f_Q}
    f_Q = m n Q^{n-1}~~~\text{and}
\end{equation}
\begin{equation}\label{Model I f_QQ}
    f_{QQ} = mn(n-1) Q^{n-2}~~~.
\end{equation}
Setting the values of $f_Q$ for the specific model-I into the equation \eqref{expression of Q}, we obtain the additional constraints
\begin{equation}
    z=\frac{2 n (x-1)+1}{1-2 n} = 1 - \frac{2n x }{2n-1}~~.
\end{equation}
Substituting equations \eqref{Model I f_Q} and \eqref{Model I f_QQ} in equation \eqref{Gamma equation}, we obtain,
\begin{equation}\label{Model I Gamma equation}
    \Gamma =\frac{1}{n-1}~~.
\end{equation}
Substituting the expression of $\Gamma$ from \eqref{Model I Gamma equation}, we get the dynamical system as,
\begin{equation}
    x^\prime = 3 x \left\{{\frac{
   2 x (1 + \omega)}{(2 + \frac{1}{n-1}) (z-1)} + 
   2 (x + z + x \omega)}-1-\omega\right\}~~~ \text{and}
\end{equation}
\begin{equation}
    z^\prime = 6 z (x \omega +x+z-1)~~~.
\end{equation}
For this particular model, $f(Q) = mQ^n$, equation \eqref{Reduced H equation} gives,
\begin{equation}
    \frac{\dot{H}}{H^2} = \frac{3 z \left\lbrace{z-1-2 n (z-1)}\right\rbrace-3 x (\omega +1) \left\lbrace{(2 n-1) z-1)}\right\rbrace}{(2 n-1) (z-1)}~~~.
\end{equation}

For the de Sitter fixed point identified in this model, the autonomous system yields $\dot H=0$, implying
\begin{equation}
\omega_{\mathrm{eff}} = -1 ,
\end{equation}
which confirms its interpretation as an inflationary or dark--energy--dominated solution. 

In addition, near the non-hyperbolic fixed point associated with the shear degree of freedom, the centre manifold analysis shows that $\dot H \ll H^{2}$, with $\dot H$ governed by higher order terms proportional to $(n-1)^{2}$. Consequently,
\begin{equation}
\omega_{\mathrm{eff}} \simeq -1 + \mathcal{O}\!\left((n-1)^{2}\right),
\end{equation}
indicating a phase that closely mimics ultra slow roll expansion.

See Appendix-II (B.1) for a brief summary of derived expressions.

\begin{table}[h!]
    \centering
    \begin{tabular}{|>{\centering\arraybackslash}m{1.5cm}|>{\centering\arraybackslash}m{1.5cm}|>{\centering\arraybackslash}m{2.2cm}|>{\centering\arraybackslash}m{2.5cm}|>{\centering\arraybackslash}m{3.5cm}|>{\centering\arraybackslash}m{1.2cm}|>{\centering\arraybackslash}m{2.4cm}|}
    \hline
      {\bf Crit.} & {\bf Exist-} & {\bf Eigenvalues} & {\bf Type of crit.} & {\bf Stability} & {\bf $C^2_{T}$} & {\bf Cosmology} \\
      {\bf Point} & {\bf ence} & & {\bf point} & & &\\
       \hline
       $x\to 0$ & $\forall \omega$ & $-6$ and $-6$ & Hyperbolic  & Stable node & & $a(t)=a_0 e^{H_0 t}$ \\
       &&& $(\forall \omega)$ && & (i.e., de Sitter) \\
       \cline{1-5}\cline{7-7}
       $x\to \frac{2 n-1}{2 n}$ & $n \neq 0$, & $\frac{3}{n}\left\lbrace{(2n-1) \omega}\right.$  & Hyperbolic  & Stable node for & & \\
       & $\forall \omega$ & $\left.{ -1}\right\rbrace$ and & $\left(\omega \neq -1,\frac{1}{2n-1}\right)$ & $(\omega<-1)$ & &\\
       \cline{5-5}
       & & $3 (\omega +1)$ & & Saddle point for & $\frac{1}{2n-1}$ &  \\
       &&&& $\left(\omega>-1,~\omega <\frac{1}{2n-1}\right)$ & for $n>0$ & $ a(t) \propto t^{-\frac{2n}{3 (\omega+1)}}$\\
       \cline{5-5}
       &&&& Unstable node for & & \\
       &&&& $\left(\omega >-1,~\omega>\frac{1}{2n-1}\right)$ & &\\
        \cline{4-5}
       &&& Non-hyperbolic  & Central manifold for & &\\
       &&&$\left(\omega =-1, \frac{1}{2n-1}\right)$& $\left(\omega=-1, \frac{1}{2n-1}\right)$ & &\\
       \hline
    \end{tabular}
    \caption{Critical points with existence conditions, eigen values, type of critical points, stability, cosmology pattern and value of $C^2_{T}$ for model-I: $f(Q) = mQ^n$ (see Appendix-I (A.1) for detailed calculation).}
\end{table}

\begin{figure}[h!]
\begin{center}
\subfloat[] {\includegraphics[height=3in,width=3in]{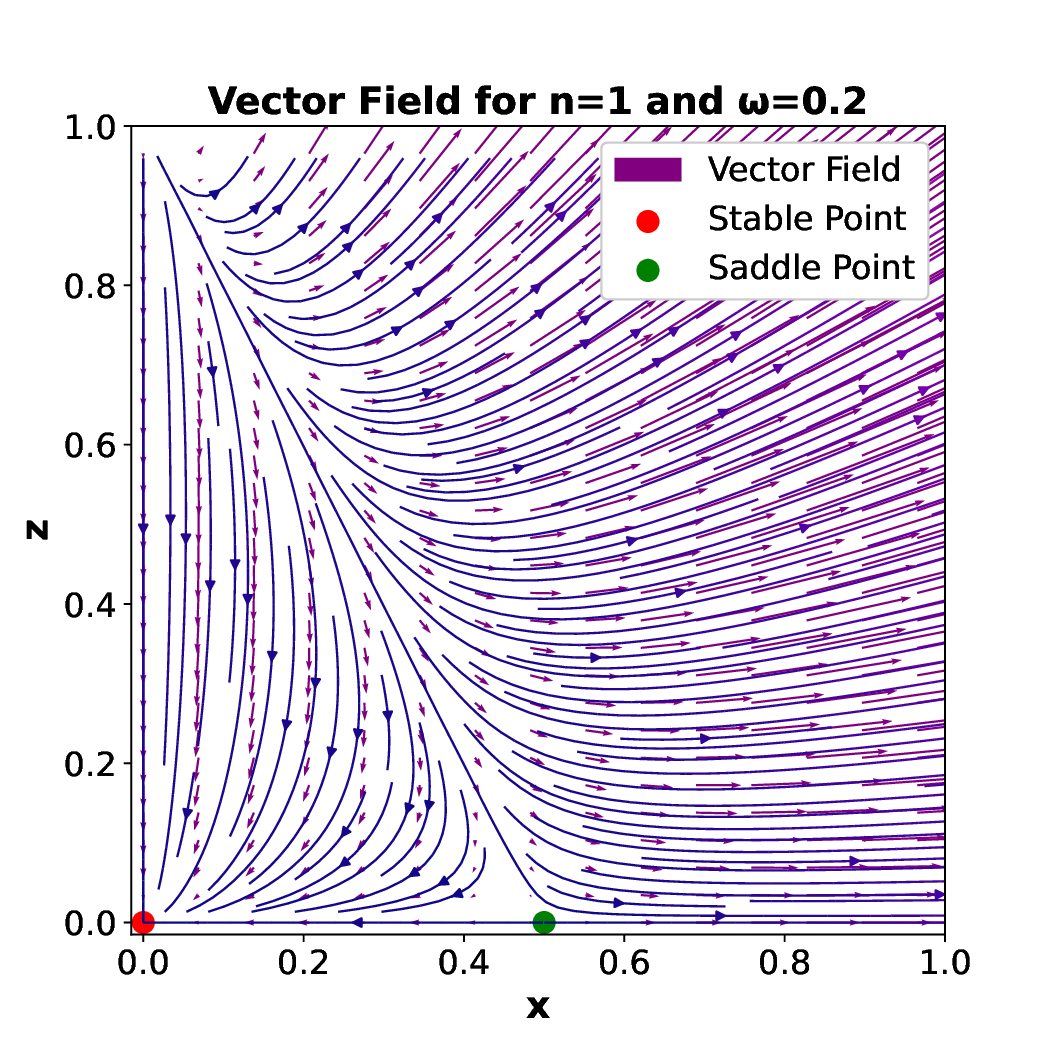}}\hspace{1cm}
\subfloat[] {\includegraphics[height=3in,width=3in]{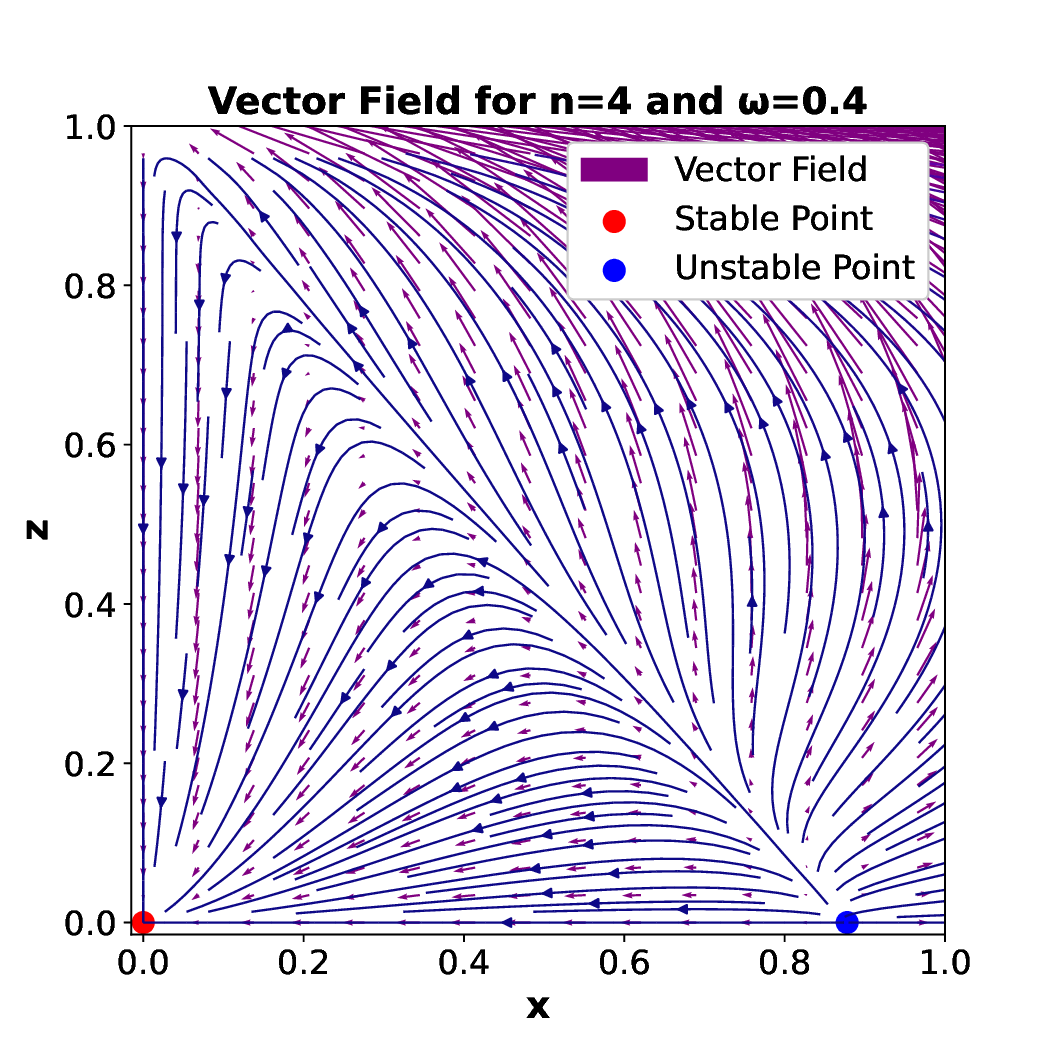}}\\
\subfloat[] {\includegraphics[height=3in,width=3in]{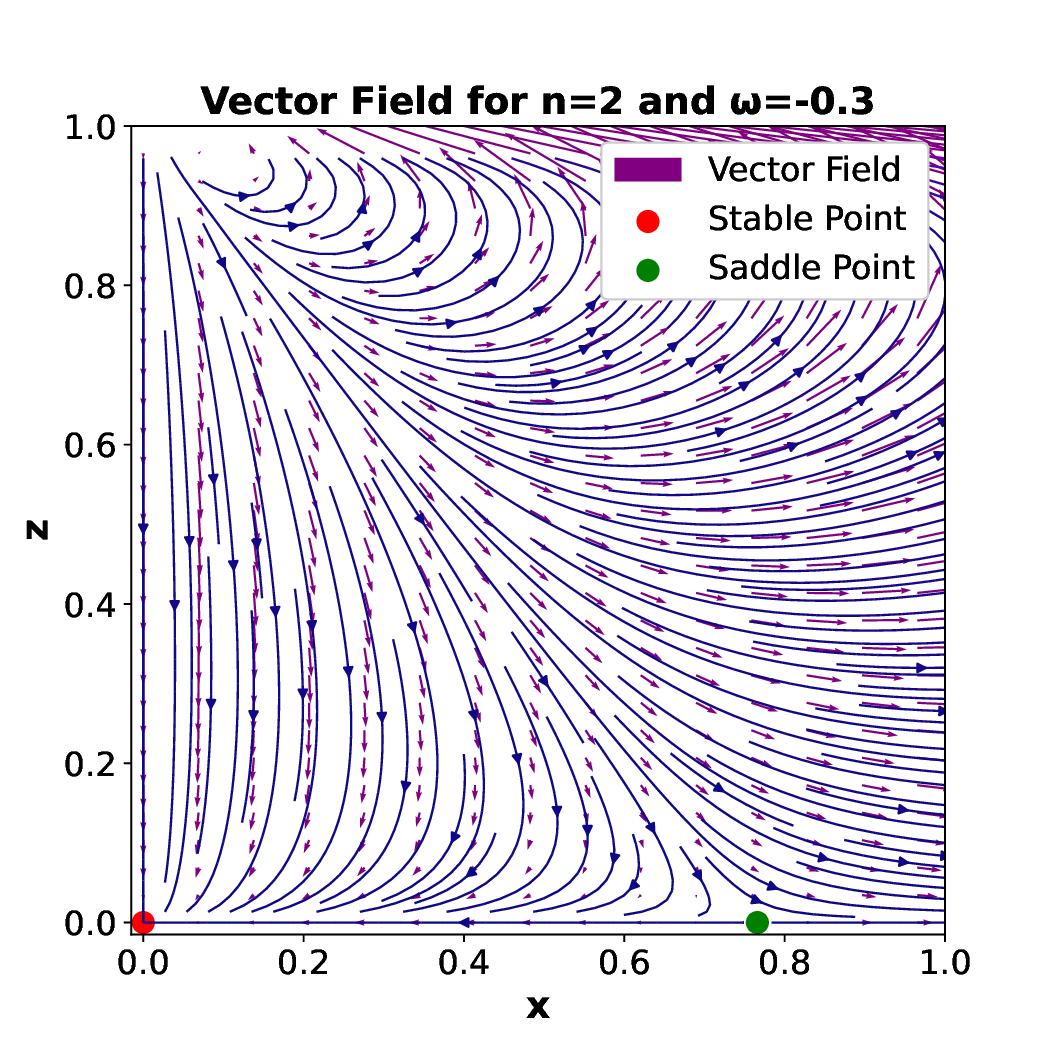}}\hspace{1cm}
\subfloat[] {\includegraphics[height=3in,width=3in]{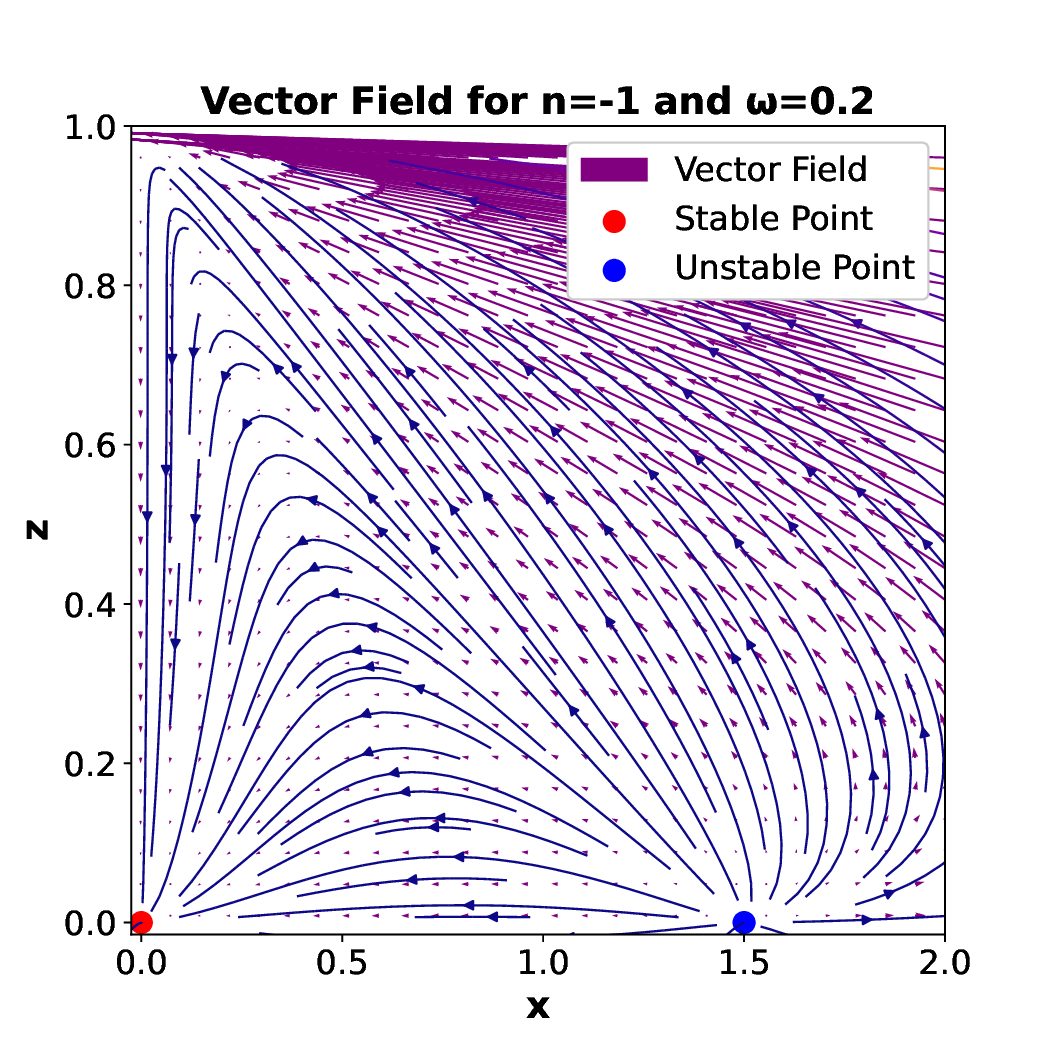}}\\
\caption{Figure $(a)-(d)$ are $z$ vs $x$ phase diagrams for the model $f(Q) = mQ^n$ in Bianchi-I cosmology at different values of $n$ and $\omega$. A red dot indicates a stable point, a green dot indicates a saddle point and a blue dot indicates an unstable point.}
\end{center}
\end{figure}

In table 1, we note all the categories of critical points. Fig $1a-1d$ are $z$ vs $x$ phase diagrams for different values of $n$ and $\omega$. Red, green and blue dots do signify stable, saddle and unstable points respectively.

Hyperbolic stable points ensure the universe always ends up in the same state, regardless of tiny differences in the past. This is important for explaining why today's universe is so isotropic and accelerating. As the hyperbolic stable point lies on the axis, the universe naturally evolves to an isotropic state. Again the hyperbolic stable point corresponds to $x=0$, i.e., $\dot{H}=0$, it gives constant Hubble rate, i.e., accelerating expansion like DE. Only certain values of $n$ in $f(Q)=mQ^n$ give realistic cosmic evolution with a hyperbolic stable attractor. Now to explain hyperbolic saddle, we notice that the saddle points with non-zero shear $(E^2 \neq 0)$ are often formed in Bianchi-I models. These saddle points describe a universe that starts anisotropic but gradually evolves towards isotropy. Physically, this supports the idea that anisotropies decay over time due to the dynamics of $f(Q)$ gravity.

Saddle points allow the universe to pass through realistic intermediate phases, like matter dominated phase(galaxy formation) or radiation dominance (CMB decoupling). This makes $f(Q)=mQ^n$ gravity cosmologically viable, as it reproduces necessary eras of standard cosmology before reaching a late-time attractor (accelerated expansion). In Bianchi-I cosmology, if a saddle point has non-zero anisotropy, the system may start near this anisotropic state and then exits toward an isotropic attractor \cite{Mandal2020}.

A hyperbolic unstable point is a natural candidate for the early universe. This represents a Big Bang like state (high curvature $Q$, large $H$). Universe emerges near this state and then evolves away deterministically. Since it is hyperbolic and structurally stable, small perturbations would not change the qualitative behavior. Again Bianchi-I type early universe can be predicted which evolves quickly away, often into a matter saddle point and finally into a de Sitter attraction. No fine-tuning even is needed as the unstable node or foci repels trajectories generally and hence a wide range of initial conditions evolve naturally form this early phase \cite{Mandal2020}.

Figure $1c$ spots a centre however. This points towards an oscillatory cosmology or a bouncing universe. Also, in a Bianchi-I geometry with shear, centres may correspond to oscillations in anisotropy. The shear variables oscillate around a mean, while the universe may slowed expand or remains bounded. These cycles can resemble mixmaster-like dynamics or approaches very slowly towards a stable de Sitter or be chaotic but bounded states \cite{Bahamonde2023}.

For the critical point $\left(\frac{2n-1}{2n},0\right)$, $z$ is the stable variable and $x$ is the central variable. Let the center manifold theory has the form $z=h(x)$, then the approximation $N$ is given as,

\begin{align*}
N(h(x))=h^\prime (x) \cdot x^\prime - z^\prime = h^\prime (x) \cdot3 x \left[{ \frac{2 x (1 + \omega)}{\left(2 + \frac{1}{n-1}\right) \{h(x)-1\}} + 
   2 \{x + h(x) + x \omega\}-1 - \omega}\right]\\- [6 h(x) \{(x \omega +x+h(x)-1\}]~~.
\end{align*}

For the zeroth approximation, the component turns to
$$N(h(x))= 0 + \mathcal{O} (x^2)~~.$$

To clarify the nature of the non-hyperbolic fixed points, we perform a minimal centre manifold analysis for the de Sitter type solution in the power-law model $f(Q)=m Q^{n}$. Linearising the system around the fixed point $(x_\ast,y_\ast,z_\ast)=(x_\ast,y_\ast,0)$ yields one vanishing eigenvalue, associated with the shear variable $z$, while the remaining eigenvalues have negative real parts.

Introducing perturbations $u=z$ and $v=(x-x_\ast,y-y_\ast)$, the system can be written in centre stable form as $\dot u = F(u,v)$ and $\dot v = A v + G(u,v)$, where the eigenvalues of $A$ have negative real parts and $F,G$ start at quadratic order. The centre manifold is locally defined by $v=h(u)$ with $h(0)=h'(0)=0$, implying $h(u)=\mathcal{O}(u^2)$.

Using the shear evolution equation in Bianchi-I $f(Q)$ cosmology, the reduced dynamics on the centre manifold takes the form
\begin{equation}
\dot u = \gamma u^{2} + \mathcal{O}(u^{3}),
\end{equation}
where $\gamma \propto (n-1)$. This shows that the evolution along the centre direction is slow and governed by nonlinear terms, confirming that the fixed point is non-hyperbolic, with stability determined by the sign of $\gamma$.

Introducing scalar perturbation to the metric and matter fields, we find
\begin{equation}
    ds^2 = -(1+2\Phi)dt^2+a_{i}^2(1-2\psi)dx^idx^i ~~~,
\end{equation}
\begin{equation}
    \rho=  \bar{\rho} + \delta\rho ~~~\text{and}~~~p = \bar{p}+\delta p~~~,
\end{equation}
assuming $\psi_1=\psi_2=\psi_3 = \psi$ (isotropic perturbations), the perturbed field equations are studied : 

Time-time perturbation for this model is given by,
\begin{equation}
    6mn(2n-1)H^2(\dot{\psi}+H\phi)+mn(n-1)Q^{n-2}\delta Q=\delta \rho   ~~~\text{and}
\end{equation}
space-space perturbation takes the form of,
\begin{equation}
    2mn Q^{n-1}(\dot{\psi}+H\phi)+mn(n-1)Q^{n-2}(\dot{Q}\psi+Q\dot{\psi} )=-a^2\delta p~~,
\end{equation}
where the non-metricity perturbation $\delta Q$ is given by, 
\begin{equation}
    \delta Q = 12H(\dot{\psi}+H\phi)~~~~.
\end{equation}
For the stability, sound speed squared $C_{s}^2 = \frac{\delta p}{\delta {\rho}}$ must be real and positive. For the model $f(Q) = mQ^n$, the conditions are,
\begin{equation}
    C_s^2= \frac{2n-1}{3n-1} ~~~~ and ~~~~mnQ^{n-1}>0.
\end{equation}

We emphasise that, although the background spacetime is anisotropic (Bianchi-I), the scalar perturbation sector in the present analysis is treated using an isotropised ansatz, namely $\psi_1=\psi_2=\psi_3$. This assumption effectively reduces the scalar perturbations to an isotropic form, allowing us to obtain first insights into stability properties while keeping the analysis tractable.

Within this simplified perturbation scheme, we define the scalar sound speed as
\begin{equation}
c_s^2 \equiv \frac{\delta p}{\delta \rho},
\end{equation}
which should be interpreted as a useful diagnostic for the behaviour of the fluid--scalar sector rather than as a complete stability criterion for all scalar degrees of freedom in $f(Q)$ gravity. In particular, this definition does not capture the full propagation properties of potential additional scalar modes arising from the modified gravity sector, which would require a systematic derivation from the quadratic action.

Thus for $n> \frac{1}{2}$, $c_s^2>0$ and $m>0$, the no-ghost condition holds if $Q>0$.
\begin{equation}
    \ddot{h}_{ij} + \left({3 H + (n-1) \frac{\dot{Q}}{Q}}\right) \dot{h}_{ij} + \frac{k^2}{a^2} h_{ij} =0~~.
\end{equation}

A fully general perturbation analysis on a Bianchi-I background, including genuinely anisotropic scalar perturbations and a derivation of the canonical scalar modes from the quadratic action, lies beyond the scope of the present work. We therefore regard the scalar perturbation results reported here as necessary but not sufficient conditions for theoretical viability, and leave a more complete anisotropic treatment for future investigation.

While working with the tensor perturbation, the tensor modes propagate with the speed :
\begin{equation}
    C_T^2 =\frac {f_Q}{f_Q+2Qf_{QQ}}= \frac{1}{2n-1} ~~.
\end{equation}

The expression for the tensor propagation speed,
\begin{equation}
c_T^2 = \frac{f_Q}{f_Q + 2Q f_{QQ}},
\end{equation}
is obtained within the perturbative framework adopted here for a Bianchi-I background and may be viewed as an extension of the corresponding FLRW results under specific assumptions. We note that, in the broader $f(Q)$ literature, gravitational waves are often found to propagate luminally under fairly general conditions, and the precise form of $c_T^2$ can depend on the background geometry and perturbation variables employed.

Hence, $C_T^2$ is independent of $m$ and $Q$ (as the background cancels out). Now, $C_T^2 =1$ iff $2n-1=1 ~~~\implies 2n-1=1~~~\implies n= 1$ and for positivity, (i.e., no gradient instability) we require that $C_T^2>0~~~\implies 2n-1>0 ~~~\implies n >\frac{1}{2}$.

We emphasise that the expressions for the tensor propagation speed $c_T^2$ employed here are obtained by extending the well-established FLRW perturbation results in $f(Q)$ gravity to the anisotropic Bianchi-I background. In a strongly anisotropic spacetime, the decomposition into pure tensor, vector, and scalar modes is more subtle, and a fully rigorous treatment would require a dedicated perturbation analysis beyond the scope of the present work.

Our expressions for $c_T^2$ are consistent with those obtained in the general perturbative analyses of $f(Q)$ gravity in isotropic backgrounds, as discussed in several recent works. In this sense, the conditions $c_T^2 \simeq 1$ and $c_s^2 > 0$ should be regarded as necessary (but not sufficient) viability criteria when applied to anisotropic cosmologies.

If one allows for $c_T^2\neq1$ at late-times, stringent observational constraints from multi-messenger events such as GW170817/GRB170817A require the gravitational wave speed to be extremely close to unity today. In the context of the power-law model $f(Q)\propto Q^n$, this implies that the parameter $n$ must be very close to $1$ at late-times, significantly restricting deviations from General Relativity. Consequently, values of $n\neq1$ should be interpreted with caution and may be more relevant for early-time or intermediate cosmological dynamics rather than for the present Universe.

GW170817 type bound (which constrain $\left\vert{C_T-1}\right\vert$ to be $\mathcal{O}\left({10^{-15}}\right)$ at low redshift) implies $\left\vert{n-1}\right\vert \lesssim 10^{-15}$. So the value of $n$ must be equal to 1 for extraordinary precision.

In the power-law model $f(Q) \propto Q^n$, the relation $c_T^2 = 1/(2n-1)$ implies that the gravitational-wave--wave constraint from GW170817 enforces $n$ to be extremely close to unity at late-times. This significantly restricts the viability of deviations from General Relativity in the present Universe. Consequently, any cosmological role of $n \neq 1$ must either be confined to early-time dynamics or interpreted within scenarios where the effective theory evolves toward $n \to 1$ at late-times.

Overall, we regard the perturbative analysis presented here as a first exploratory step toward a fully consistent anisotropic perturbation framework in $f(Q)$ gravity. By making the underlying assumptions and limitations explicit, we aim to provide a transparent basis for future studies incorporating a complete anisotropic scalar sector and a systematic effective field theory treatment.

In equation of $x^{\prime}$, we can substitute the value of $z$ and obtain,
\begin{equation}
    x^\prime = \frac{3\{1+2n(x-1)\}x\{(2n-1)\omega-1\}}{n(2n-1)}    ~~~~ \text{and} 
\end{equation}
in equation of $z^\prime$, substituting the value of $x$, we get,
\begin{equation}
    z^\prime= -\frac{3}{n} (z-1)z\{(2n-1)\omega-1\}   ~~~~~~~~\text{[by incorporating the value of $z$]}~~~~.
\end{equation}
Linearizing around $x=0$, we set $x=\delta x$ and obtain, 
\begin{equation}
    \frac{d(\delta x)}{dN} \approx\frac{3(1-2n)\{(2n-1)\omega-1\}\delta x}{n(2n-1)}=-\frac{3}{n}\{(2n-1)\omega -1\}\delta x
\end{equation}
\begin{equation}
    \Rightarrow \delta x \sim exp\left\{{-\frac{3}{n}\{(2n-1)\omega-1\}N}\right\}
\end{equation}
Similarly, linearizing around $z=0$, we set $z=\delta z$ to follow that. Then, 
\begin{equation}
    \frac{d(\delta z)}{dN}\approx\frac{3}{n}\{(2n-1)\omega-1\}\delta z ~~~ \Rightarrow \delta z \sim exp\left\{{\frac{3}{n} \left\lbrace{(2n-1) \omega -1}\right\rbrace N}\right\}~~~.
\end{equation}

At last scattering (when the CMB was emitted) photons left the surface at the same temperature. But if the universe expands anisotropically, then photon along faster expanding directions get more redshifted, arriving cooler today and along slower expanding directions get less redshifted arriving hotter. This directional differences produces a large scale temperature pattern (primarily a quadrupole)
\begin{equation}
    \frac{\Delta T}{T}(\hat{n}) \approx - \int_{t_{1s}}^{t_0} \sigma_{ij} n^i n^j dt~~~,
\end{equation}
where $\frac{\Delta T}{T}$ is the CMB temperature anisotropy, $\hat{n}$ is the direction along which CMB is measured in the sky, $\sigma_{ij}$ is the shear tensor, $t$ is the time and $n^i$ is the photon's direction. From this, we say that the anisotropy is proportional to the integrated shear between last scattering and today.

The amplitude of this quadrupole roughly scatters as, $\left(\frac{\Delta T}{T}\right)_{quad} \sim \frac{\sigma}{H}$, where $\sigma$ is the shear scalar and $H$ is the Hubble parameter. Using the values of Planck's observed quadrupole for an anisotropic expansion, we get $\left(\frac{\sigma}{H}\right)_{0} \simeq 5 \times 10^{-11}$, where $( )_0$ denotes the value at present (i.e., at redshift 0).

For the model-I and a Bianchi-I space-time, the field equations may lead to the form
$$\dot{\sigma}_{ij}+ 3H\xi(n)\sigma_{ij}=0~~~,$$
where $\xi(n)$ depends on $n$ and given by $\xi(n)=1 + \frac{n-1}{Q} \frac{\dot{Q}}{3H}$.

If $\xi(n)<1$, the shear decays more slowly, possibly giving residual anisotropy that obey the Planck constraint above.

%If we follow matter domination, for pure matter $\rho=3H^2$, i.e., $x=1$ and hence $\frac{dx}{dN}=0$. If we linearize near $x=1$, assuming $\frac{dx}{dN}=f(x)$, we can obtain $f'(1)=\frac{3}{n}\frac{(-2 n \omega +\omega +1)}{(1-2 n)}~.$

%%%%%%%%%%%%%%%%%%%%%%%%%%%%%%%%%%%%%%%%%%%%%%%%%%%%%%%%%%%%%%%%%%%%%%%%%%%%%
\subsection{Model-II : $f(Q)= exp \{nQ\}$}
%Here, $n$ is a constant parameter that controls the strength of the deviation from GR. For small $nQ$ 
%This shows that $f(Q) \approx Q$ in the limit $n\rightarrow 0$, which recovers GR. The exponential form introduces the non-linear effects that can help to explain the late-time cosmic acceleration and avoid the cosmological singularities. It also provides better fits for the cosmological data. 

In contrast to many higher order curvature based theories like $f(R)$ etc., $f(Q)$ gravity leads to second order field equations which are similar to those produced by GR. Higher order time derivatives typically signal ghost instabilities, eg. Ostrogradsky ghosts. As exponential $f(Q)$ model inherits this second order nature, it turns dynamically stable at the level of equations of motion. The first Friedmann equation becomes $e^{6nH^2}\left(6nH^2+\frac{1}{2}\right)=\rho$. At late-times, $\rho\rightarrow 0$, as matter dilutes. Now $e^{6nH^2}>0$ and $6nH^2+\frac{1}{2}\rightarrow 0$ which gives a real (positive) value of Hubble parameter as $H^2\rightarrow \frac{1}{\left|12n\right|}$. This signifies that the Hubble parameter asymptotes to a constant, corresponding to de Sitter expansion. This resembles cosmological constant. Also, the effective equation of state $\omega_{eff}=-1-\frac{2\dot{H}}{3H^2}$ turns $-1$ at late-time (as $H\rightarrow$ constant and $\dot{H}\rightarrow 0$). This also supports cosmological constant at late-time. At small $Q$, i.e., small Hubble parameter, we can expand the function as,
\begin{equation}
    f(Q) \approx 1+nQ +\frac{(nQ)^2}{2!}+ \cdots~~.
\end{equation}
This reduces to GR when $nQ<<1$, making it easy to recover GR in an appropriate limits and controls the deviation. Depending on the sign and magnitude of $n$, the model can lead to bouncing cosmologies, de Sitter solutions, phantom  divergence free evolution, emergent universe scenarios and builds an outline to non-standard cosmological histories.
 
%The exponential $f(Q)$ model was tested against various cosmological datasets, including the Pantheon+ supernova compilation, Hubble parameter measurements and redshift space distortion data. The researchers employed a Markov Chain Monte Carlo (MCMC) analysis to numerically solve the modified Friedmann equations and constrain the model parameters. Their findings indicate that the exponential model provides a statistically better fit to the observational data compared to both the power law $f(Q)$ model and the standard $\Lambda$CDM model, as evaluated using Bayesian and corrected Akaike Information Criteria\cite{Mhamdi2024}. 

For this model, we can express $f_Q$ and $ ~f_{QQ}$ in terms of the dynamical variables as,
\begin{equation}
    f_Q = n e^{n Q}~~~\text{and}
\end{equation}
\begin{equation}
    f_{QQ} = n^2 e^{n Q}~~~.
\end{equation}
Substituting the value of $f_Q$ in \eqref{expression of Q}, we obtain,
\begin{equation}
    Q = \frac{1}{n} \cdot \frac{1-z}{2(1-x-z)}~~~~.
\end{equation}
Now we can express the $Q$-derivatives of $f(Q)$ in terms of the dynamical variables as,
\begin{equation}\label{Model II f_Q}
    f_Q = n~ exp\left\{\frac{1-z}{2(1-x-z)} \right\}~~~\text{and}
\end{equation}
\begin{equation}\label{Model II f_QQ}
    f_{QQ} = n^2~ exp\left\{\frac{1-z}{2(1-x-z)} \right\}~~~~.
\end{equation}
Substituting the values of \eqref{Model II f_Q} and \eqref{Model II f_QQ} in \eqref{Gamma equation}, we obtain, 
\begin{equation}\label{Gamma Model II}
    \Gamma = \frac{f_Q}{Q f_{QQ}}= \frac{1}{nQ}=\frac{2(1-x-z)}{1-z}~~~.
\end{equation}
Now by substituting the expression of $\Gamma$ from \eqref{Gamma Model II}, the dynamical equations reduced to,
\begin{equation}
    x^\prime = \frac{6 x \left\{x^2 (\omega +1)-2 x (\omega +1)+x (2 \omega +3) z+(z-1) (2 z-1-\omega)\right\}}{x+2 z-2}~~~\text{and}
\end{equation}
\begin{equation}
    z^\prime = 6 z (x \omega +x+z-1)~~~.
\end{equation}
For this particular model, the equation \eqref{Reduced H equation} is reduced to,
\begin{equation}
    \frac{\dot{H}}{H^2} = -\frac{3 \left[x^2 (\omega +1)+x \left\{(2 \omega +3) z-1-\omega \right\}+2 (z-1) z\right]}{x+2 z-2}~~~.
\end{equation}

This model admits a saddle point corresponding to a matter dominated epoch, for which the effective dynamics yields $\dot H \simeq -\tfrac{3}{2}H^{2}$ and hence
\begin{equation}
\omega_{\mathrm{eff}} \simeq 0 .
\end{equation}
At late-times, when the nonlinear term dominates, the system evolves toward an accelerating attractor with $\dot H \rightarrow 0$, leading to
\begin{equation}
\omega_{\mathrm{eff}} \rightarrow -1 ,
\end{equation}
consistent with a dark energy dominated phase.

See Appendix-II (B.2) for a brief summary of derived expressions.

\begin{table}[h!]
    \centering
    \begin{tabular}{|>{\centering\arraybackslash}m{1.5cm}|>{\centering\arraybackslash}m{1.5cm}|>{\centering\arraybackslash}m{2.2cm}|>{\centering\arraybackslash}m{2.5cm}|>{\centering\arraybackslash}m{3.5cm}|>{\centering\arraybackslash}m{1.3cm}|>{\centering\arraybackslash}m{2.4cm}|}
    \hline
       {\bf Crit.} & {\bf Exist-} & {\bf Eigenvalues} & {\bf Type of crit.} & {\bf Stability}  & {\bf $C^2_{T}$} & {\bf Cosmology}\\
       {\bf Point} & {\bf ence} & & {\bf point} & & &\\
       \hline
       $x\to 0$ & $\forall \omega$ & $-3 (\omega +1)$ &  Hyperbolic & Stable node for & & $a(t)=a_0 e^{H_0 t}$\\
       & & and $-6$ & $(\omega \neq -1)$ & $\omega>-1$ & & (i.e., de Sitter) \\
       \cline{5-5}
       & & &  & Saddle Point for & &\\
       &&&& $\omega<-1$ & $\frac{1}{1+2nQ}$ & \\
       \cline{4-5}
       & & & Non-hyperbolic & Center Manifold & with &  \\
       &&&  $(\omega=-1)$ &  for $\omega=-1$ &  $2nQ<<1$ & \\
       \cline{1-5}\cline{7-7}
       $x\to 1$& $\forall \omega$ & $0$ and $6 \omega$ & Non-hyperbolic & Fully degenerate  & & $a(t)=a_0 e^{H_0 t}$ \\
       
       &&& $(\forall \omega)$ & for $\omega =0$ & & (i.e., de Sitter) \\
       \cline{5-5}
       & & & & unstable for & & \\
       & & & & $\omega>0$ & & \\
        \cline{5-5}
       & & & & Center Manifold & & \\
       &&&& for $\omega <0$ & &\\
       \hline
    \end{tabular}
\caption{Critical points with existence conditions, eigenvalues, type of critical points, stability, cosmology pattern and value of $C^2_{T}$ for model-II: $f(Q)= exp \{nQ\}$ (see Appendix-I (A.2) for detailed calculation).}
\end{table}

\begin{figure}[h!]
\begin{center}
\subfloat[] {\includegraphics[height=3in,width=3in]{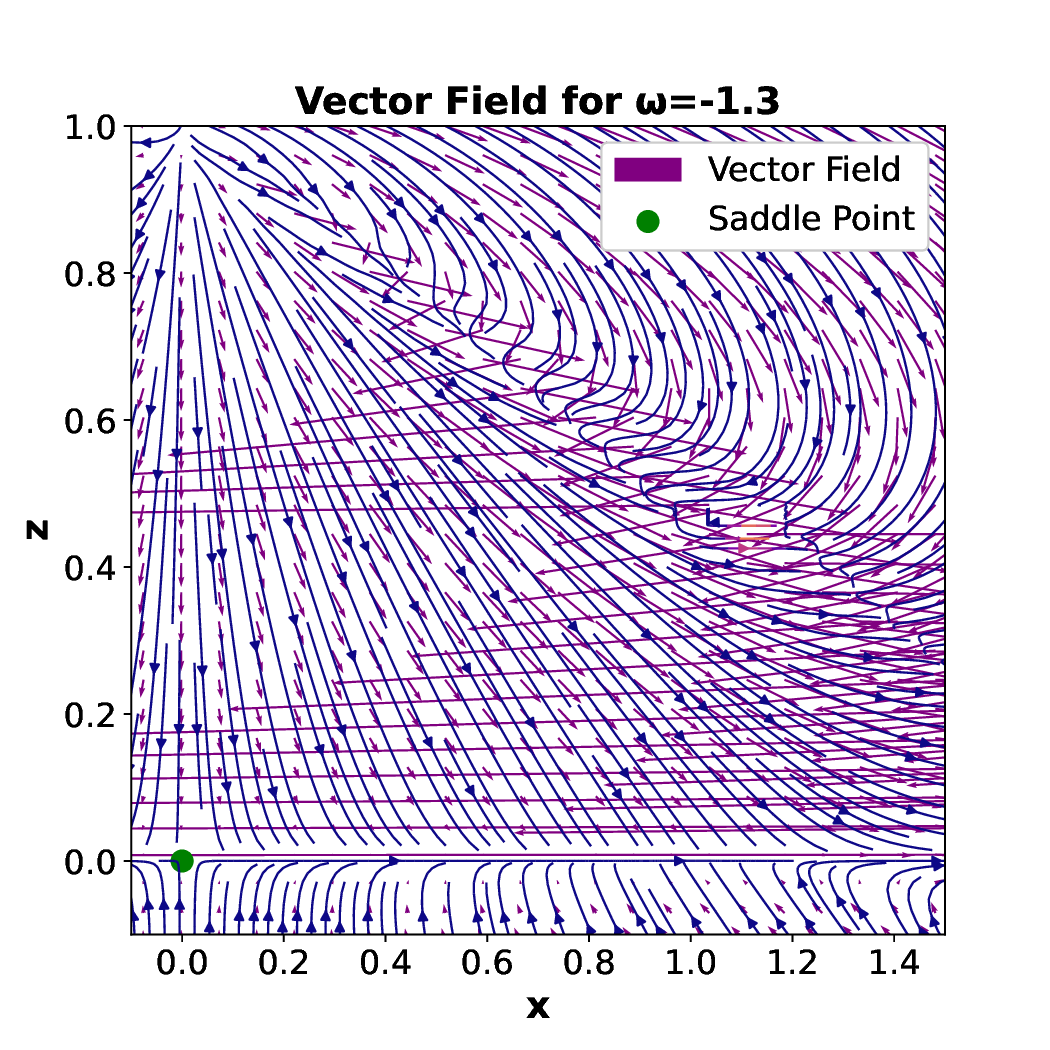}}\hspace{1cm}
\subfloat[] {\includegraphics[height=3in,width=3in]{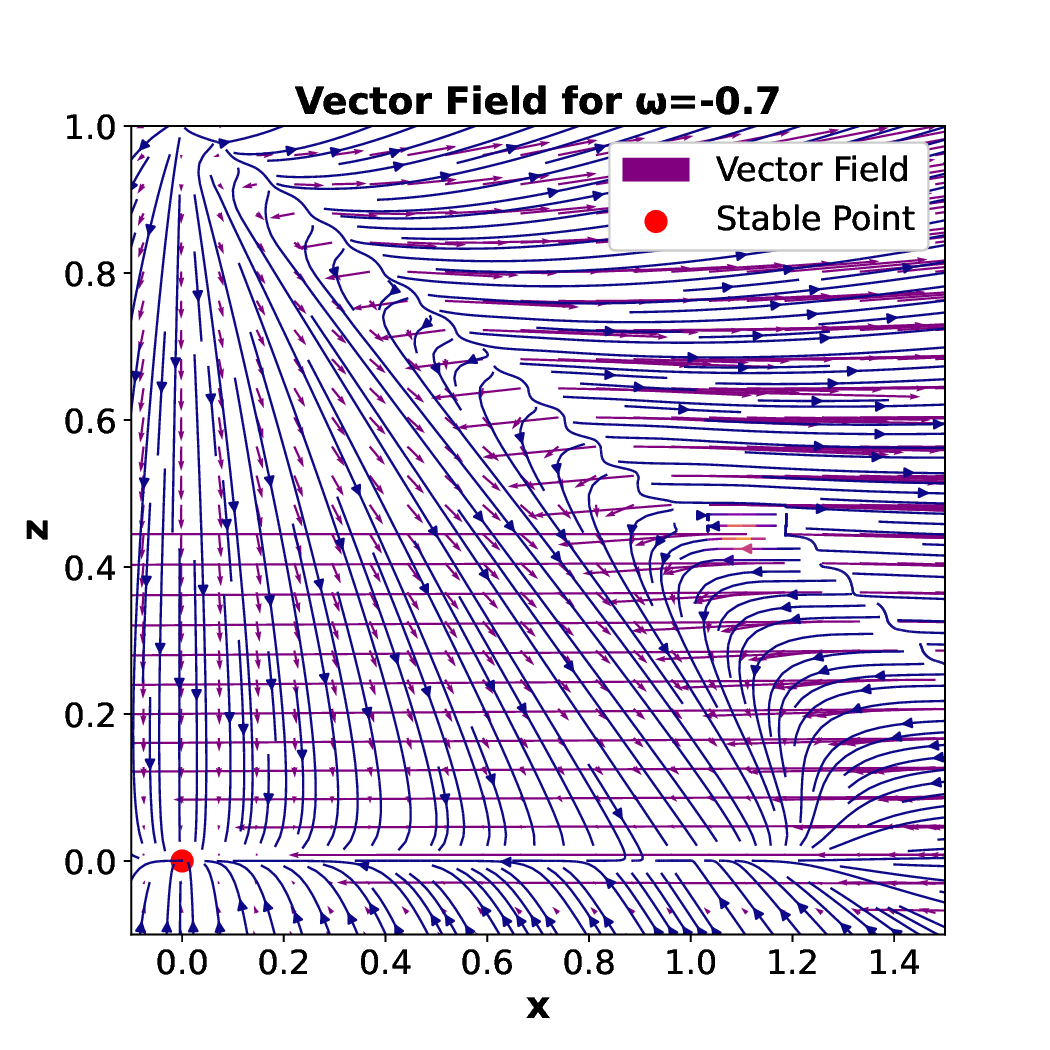}}\\
\subfloat[] {\includegraphics[height=3in,width=3in]{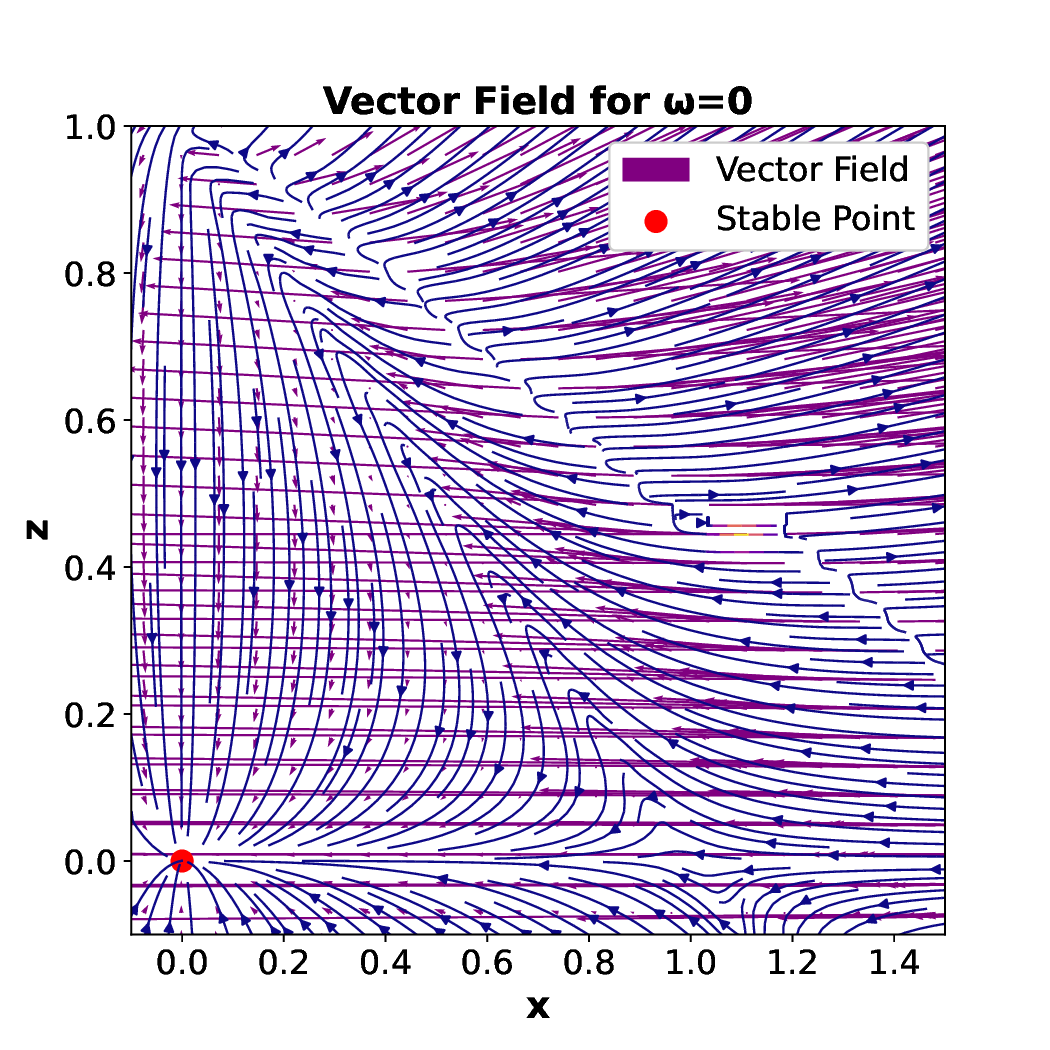}}\hspace{1cm}
\subfloat[] {\includegraphics[height=3in,width=3in]{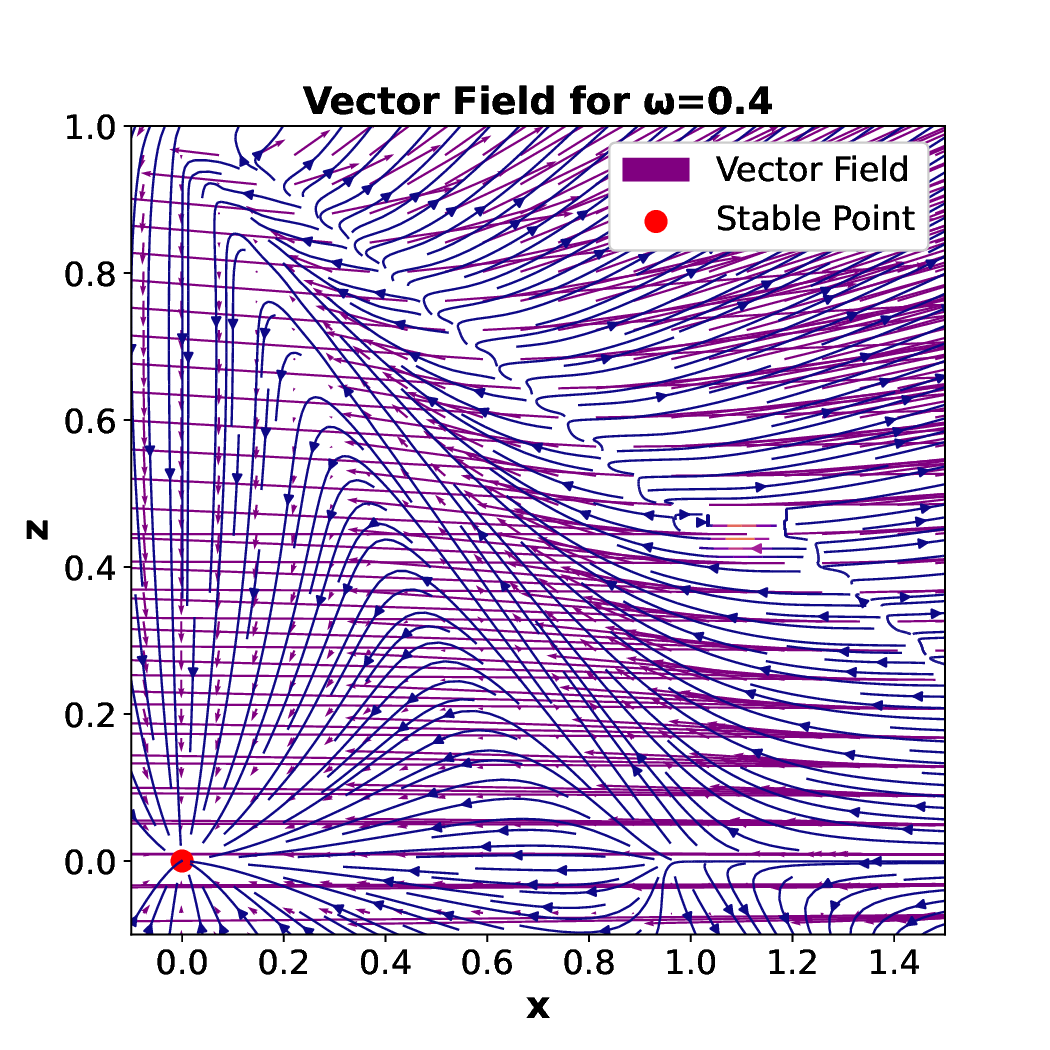}}\\
\caption{Figure $(a)-(d)$ are $z$ vs $x$ phase diagrams for the model $f(Q)= exp \{nQ\}$ at different values of $\omega$. A red dot indicates a stable point, a green dot indicates a saddle point and a blue dot indicates an unstable point.}
\end{center}
\end{figure}
Different critical points and their categorizations are given in the table 2. In fig $2a$ to $2d$, $z$ vs $x$ phase diagrams for different values of $\omega$ are drawn. 

Stable hyperbolic nodes are followed which corresponds to a de Sitter like state where scale factor grows exponentially $(\approx e^{H(z)})$. This aligns with attempts to explain cosmic acceleration via pure gravity modifications. Matter energy density $\Omega \to 0$ at stable nodes \cite{Frusciante2022}. Anisotropy variables decay and the universe isotropizes and becomes matter diluted at late-times. Formation of such node improves the predictive stability of the theory across cosmological scales.

Hyperbolic saddle generally denotes matter or radiation dominated phases. $\Omega_m$ is dominant in this case and the shear may decay slowly. These epochs are required for structure formation and CMB consistency. Saddles ensure that the universe passes through standard observable stages like matter domination, before exiting towards a DE dominated future. In $f(Q)= exp\{nQ\}$, the exponential term gradually becomes dominant at low curvature $(Q\rightarrow 0)$. Saddle behavior allows the system to stay long enough in matter dominated phase and then naturally exit towards a stable attractor, typically which is a de Sitter node. This provides a graceful exit mechanism to late-time acceleration without fine-tuning. In an anisotropic Bianchi-I model, a saddle point may include non-zero shear\cite{Iozzo2021}. The eigenvalues control whether anisotropy decays (via negative eigenvalues) to isotropization or grows to instability. If shear decays, the saddle supports early anisotropic evolution that transits to isotropy $-$ consistent with observational constraints.

Saddle points connect unstable early time nodes( like Big Bang) to stable late-time de Sitter like attractors. These fixed points act as a cosmic way point that funnels trajectories toward the realistic evolution path.

In table 2, we also observe that the possibility of formation of a center manifold, when $\omega=-1$. In cosmology, generally, this corresponds to slow, delicate evolution where the universe may linger near such a regime and small changes in parameters can lead to dramatically different outcomes. The center manifold typically hosts trajectories that slowly evolve, remaining close to a fixed point for extended periods. Cosmologically, a quasi-de Sitter state resembles ultra-slow roll inflation or early time acceleration \cite{Prasia2016}. A center manifold indicates that the system is at the edge of bifurcation. Small changes in initial conditions or parameters could lead to a stable de Sitter future or collapse or oscillatory/cyclic behavior or to exit towards anisotropic Bianchi-I type attractors. In Bianchi-I models, the center manifold can host solutions where shear slowly decays or remains nearly constant. It models a universe that becomes more homogeneous gradually due to geometry.

When $x\rightarrow 1$ and $\omega=0$, we find the existence of a fully degenerate fixed point. The Jacobian matrix turns nilpotent at this point. It is non-hyperbolic and linear stability analysis breaks down. Typically, this signals the onset of a bifurcation, a phase transition or a critical behavior. Cosmologically, a sensitive region is signified where small changes in parameters generate bifurcation points where new attractor solutions appear or existing ones disappear. These attractors can be de Sitter, power law inflation, cyclic or other exotic geometries. A geometric realisation of non-singular cosmological models without requiring extra matter content is followed from such fixed points. Even this may model an ultra slow roll inflationary epoch or a prolonged quasi de Sitter phase, naturally arising from the gravitational dynamics.

For this model, $z$ is the stable variable and the central variable is $x$. Let the center manifold theory have the form $y=h_1(x)$ and $u=h_2(x)$. Then the approximation $N$ has one component:
\begin{equation}
\begin{split}
N(h(x))= h^\prime(x) \cdot x^\prime - z^\prime = h^\prime (x) \left[\frac{6 x \left\{x^2 (\omega +1)-2 x (\omega +1)+x (2 \omega +3) h(x)+(h(x)-1) (-\omega +2 h(x)-1)\right\}}{x+2 h(x)-2}\right] \\-\left[6 h(x) \{x \omega +x+h(x)-1\}\right]~~~.
\end{split}
\end{equation}

For zeroth approximation, this component turns
$$N(h(x)) = 0+ \mathcal{O}(x^2) ~~~.$$

Linearising the autonomous system around this point yields one vanishing eigenvalue and two negative eigenvalues, indicating that the fixed point is non-hyperbolic, with the zero eigenvalue associated with the shear variable $z$. To analyse the dynamics beyond linear order, we perform a minimal centre manifold reduction.

Introducing perturbations
\begin{equation}
u=z,\qquad v_1=x,\qquad v_2=y-1,
\end{equation}
the system can be written in centre stable form as
\begin{align}
\dot u &= F(u,v), \\
\dot v &= A v + G(u,v),
\end{align}
where $u$ denotes the centre variable, $v=(v_1,v_2)$ represents the stable directions, the eigenvalues of $A$ have negative real parts, and $F$ and $G$ contain only nonlinear terms.

The centre manifold is defined locally by
\begin{equation}
v = h(u), \qquad h(0)=0,\quad h'(0)=0,
\end{equation}
which implies $h(u)=\mathcal{O}(u^{2})$. Substituting this relation into the evolution equation for $u$ and expanding near $u=0$, the reduced dynamics on the centre manifold takes the form
\begin{equation}
\dot u = \gamma (m-1)\,u^{2} + \mathcal{O}(u^{3}),
\end{equation}
where $\gamma>0$ is a constant determined by the background quantities.

This result shows that the evolution along the centre direction is governed by quadratic terms, leading to slow power-law behaviour rather than exponential evolution. The stability of the fixed point along the shear direction depends on the sign of $(m-1)$: for $m<1$ the shear decays and the solution isotropises, while for $m>1$ the shear grows, rendering the fixed point unstable. In the limit $m=1$, the model reduces to the general relativistic case.

Time-time perturbation for this model emerges as
\begin{equation}
    6ne^{nQ}(\dot{\Psi}+H\Phi) + n^2e^{nQ}\delta Q = \delta \rho ~~,
\end{equation}
where $\delta Q = 12H(\dot{\Psi}+H\Phi)$.\\
Space-space perturbation, on the other hand, produces
\begin{equation}
    2ne^{nQ}(\dot{\Psi}+H\Phi) +n^2e^{nQ}(\dot{Q}{\Psi}+Q\dot{\Psi)}=-a^2\delta p ~~,
\end{equation}
where the non-metricity perturbation $\delta Q$ is, 
\begin{equation}
    \delta Q = 12H(\dot{\Psi}+H\Phi)~~.
\end{equation}
The propagation speed of scalar perturbation is,
\begin{equation}
    C_s^2 = \frac{1+n\frac{\dot{Q}}{H}}{1+2nQ} ~~ .
\end{equation}
The stability occurs when $C_s^2>0$, if $n>0$ and the no ghost condition holds if $n>0$.\\
The tensor modes propagate with the speed is,
\begin{equation}
    C_T^2 = \frac{1}{1+2nQ}~~~~.
\end{equation}
In the cosmological background of $f(Q)$ gravity, the non-metricity scalar depends on the Hubble rate by the relation $ Q=6H^2$. As the universe expands, $H(t)$ decreases dramatically ($H_{early} \sim 10^{37} ~s^{-1}$ (inflation) and $H_{0} \sim 10^{-18}~s^{-1} $). Even if $n$ has a moderate size, the product $nQ=6nH^2$ automatically tends to zero as $H\to 0$. Therefore, at the late-time attractor, $2nQ \to 0$.

For $C_T^2 \approx 1 $ (as in GR), we require $2nQ<<1$. GW constraint naturally satisfies this condition without fine-tuning. Basically, it is a built-in feature of the model's dynamical evolution.

In this sense, models with $n \neq 1$ may be viewed as phenomenological explorations of early-time or intermediate cosmological dynamics, rather than fully viable late-time alternatives without further theoretical refinement.

Under considering the Bianchi-I space-time the field equation for model-II may lead to, 
$$\dot{\sigma}_{ij}+(3H+n\dot{Q})\sigma_{ij}=0~~~.$$

This leads the solution 
$$\sigma_{ij}(t)\sim \sigma_{ij_{,}*} ~a^{-3}(t)  e^{-n\{Q(t)-Q_{*}\}}~~~,$$
where $\sigma_{ij_{,}*}$ is the initial value of the shear tensor at a chosen reference time $t_*$, $a(t)$ is the scale factor at time $t$ and $Q_*$ is the non-metricity scalar at time $t_*$.

So today,
$$\frac{\sigma_{ij}}{H}\Bigm\vert_{t_{0}}=\frac{\sigma_{ij_{,}*}}{H_{0}} a^{-3}(t_{0})  e^{-n(Q_{0}-Q_{*})}~~~,$$
where $a(t_0)$ is the scale factor at time $t_0$, $H_0$ is the Hubble constant and $Q_0$ is the present time non-metricity scalar.

The sign and magnitude of $n$ control whether shear decays faster or slower than the GR $a^{-3}$ law.\cite{saadeh2016isotropic}

%%%%%%%%%%%%%%%%%%%%%%%%%%%%%%%%%%%%%%%%%%%%%%%%%%%%%%%%%%%%%%

\subsection{Model-III : $f(Q)=\alpha Q^2+ v Q^2 \log (Q)$}
This model is inspired by quantum corrections, especially the logarithmic term arises in loop expansions \cite{BirrellDavies1982, Nojiri2004_lnR}
Effective field theories, where logarithmic corrections are expected at high curvature regimes, are also driving motivations. In the Bianchi-I metric, shear or anisotropy enters via
$$\sigma^2=\frac{1}{2}\sum_{i=1}^3(H_i-H)^2~~\text{and}$$
the non-metricity scalar $Q$ is generalised to $Q=6H^2+\text{shear ~corrections}$. At high $Q$, the $Q^2logQ$ term becomes dominant to drive inflation like dynamics. Te effective equilibrium of state $\omega_{eff}\rightarrow -1$(quasi de Sitter) and anisotropy $\sigma^2$ decays over time. When matter dominated decelerated expansion is going on or when the anisotropy slowly decays (matching CMB constraints), we expect saddle. As $Q\rightarrow 0+$, $Q^2logQ \rightarrow 0$ and hence depending on the value of $\alpha$, stable de Sitter attractor or phantom like regime ($\omega_{eff}<-1$) may be followed.

Now, for this model, we can calculate the $Q$-derivatives as,
\begin{equation}
    f_Q = (2 \alpha + v) Q+2 Q v \log (Q)~~~\text{and}
\end{equation}
\begin{equation}
    f_{QQ} = (2 \alpha+3 v) +2 v \log (Q)~~~.
\end{equation}
Substituting the values of $f_Q$ and $f_{QQ}$ in equation \eqref{expression of Q}, we get,
\begin{equation}
    \log(Q) = \frac{\alpha (3-4 x-3 z)+2 v (x+z-1)}{v (4 x+3 z-3)}~~~.
\end{equation}
Similarly, we can calculate the expression of $\Gamma$, given by,
\begin{equation}
    \Gamma = \frac{8 x+7 z-7}{16 x+13 (z-1)}~~~.
\end{equation}
Now, by substituting the values of $\Gamma,$ the dynamical variables are reduced to,
\begin{equation}
    x^\prime = 3 x \left[\frac{2 x (\omega +1) \left\{16 x+13 (z-1)\right\}}{(z-1) \left\{40 x+33 (z-1)\right\}}+2 (x \omega +x+z)-\omega -1\right]~~~\text{and}
\end{equation}
\begin{equation}
    z^\prime = 6 z (x \omega +x+z-1)~~~.
\end{equation}
Also for this model, the equation \eqref{H equation} reduces to,
\begin{equation}
    \frac{\dot{H}}{H^2} = -\frac{3 \left[8 x^2 (\omega +1) (5 z-1)+x (z-1) \left\{(33 \omega +73) z-7 (\omega +1)\right\}+33 (z-1)^2 z\right]}{(z-1) \left\{40 x+33 (z-1)\right\}}~~~~.
\end{equation}

Near the critical fixed points of this model, the evolution of the Hubble parameter can be written as $\dot H \simeq \epsilon(\alpha) H^{2}$, where $|\epsilon|\ll1$ depends on the model parameter $\alpha$. The corresponding effective equation of state is therefore
\begin{equation}
\omega_{\mathrm{eff}} = -1 - \frac{2}{3}\epsilon(\alpha).
\end{equation}
Small variations of $\alpha$ modify the sign and magnitude of $\epsilon$, leading to qualitatively different cosmological behaviour, which supports the interpretation of bifurcation like features in the phase space.

See Appendix-II (B.3) for a brief summary of derived expressions.
\begin{table}[h!]
    \centering
    \begin{tabular}{|>{\centering\arraybackslash}m{1.5cm}|>{\centering\arraybackslash}m{1.5cm}|>{\centering\arraybackslash}m{2.2cm}|>{\centering\arraybackslash}m{2.4cm}|>{\centering\arraybackslash}m{3.2cm}|>{\centering\arraybackslash}m{1.9cm}|>{\centering\arraybackslash}m{2.2cm}|}
    \hline
       {\bf Crit.} & {\bf Exist-} & {\bf Eigenvalues} & {\bf Type of crit.} & {\bf Stability} & {\bf $C^2_{T}$} & {\bf Cosmology}\\
      {\bf Point} & {\bf ence} & & {\bf point} & & & \\
       \hline
       $x\to 0, z\to 0$ & $\forall \omega$ & $-6$ and & Hyperbolic & Stable node for & & $a(t)=a_0 e^{H_0 t}$ \\
       & & $-3 (\omega +1)$ & $(\omega \neq -1)$ & $\omega >-1$ & & (i.e., de Sitter) \\
       \cline{5-5}
       & & & & Saddle Point for & & \\
       & & & & $\omega<-1$ & & \\
       \cline{4-5}
       & & & Non-hyperbolic & Center Manifold for & &\\
       & & & $(\omega =-1)$ & $\omega =-1$ & & \\
       \cline{1-5}\cline{7-7}
       $x\to \frac{3}{4}, z\to 0$ & $\forall \omega$ & $6 (\omega+1)$ and & Hyperbolic & Stable node for & & $a \propto t^{\frac{4}{3(1+\omega)}}$ \\
       & & $\frac{3}{2} \left(3 \omega -1\right)$ & $(\omega \neq -1, \frac{1}{3})$ & $\omega <-1$ & &\\
       \cline{5-5}
       && && Saddle point for & & \\
       & & & & $-1 <\omega <\frac{1}{3}$ & &\\
       \cline{5-5}
       &&&& Unstable node for & &\\
       & & & & $\omega>\frac{1}{3}$ & &\\
       \cline{4-5}
       &&& Non-hyperbolic & Unstable for & $\frac{v+2\alpha+2vlog Q}{7v+6\alpha+6vlogQ}$ &\\
       & & & $(\omega =-1, \frac{1}{3})$ & $\omega =\frac{1}{3}$ & &\\
       \cline{5-5}
       & & & & Center manifold & & \\
       & & & & for $\omega=-1$ & & \\
       \cline{1-5}\cline{7-7}
       $x\to \frac{11}{12}, z\to 0$ & $\forall \omega$ & $6 (\omega+1)$ and & Hyperbolic & Stable node for & & $a \propto t^{\frac{4}{(1+\omega)}}$\\
       & & $\frac{1}{2} \left(11 \omega -1\right)$ & $(\omega \neq -1, \frac{1}{11})$ & $\omega <-1$  & &\\
       \cline{5-5}
       &&&&Saddle point for & &\\ 
       &&&& $-1<\omega <\frac{1}{11}$ & &\\
       \cline{5-5}
       &&&& Unstable node for & &\\
       & & & & $\omega >\frac{1}{11}$ & &\\
       \cline{4-5}
       &&& Non-hyperbolic  & Unstable for & &\\
       &&& $(\omega =-1,\frac{1}{11})$ & $(\omega =\frac{1}{11})$ & &\\
       \cline{5-5}
       & & & & Center manifold & & \\
       & & & & for $\omega=-1$ & & \\
       \hline
    \end{tabular}
    \caption{Critical points with existence conditions, eigenvalues, type of critical points, stability, cosmology pattern and value of $C^2_{T}$ for model-III: $f(Q)=\alpha Q^2+ v Q^2 \log (Q)$ (see Appendix-I (A.3) for detailed calculation).}
\end{table}

\begin{figure}[h!]
\begin{center}
\subfloat[] {\includegraphics[height=3in,width=3in]{Figure/Figure_8.eps}}\hspace{1cm}
\subfloat[] {\includegraphics[height=3in,width=3in]{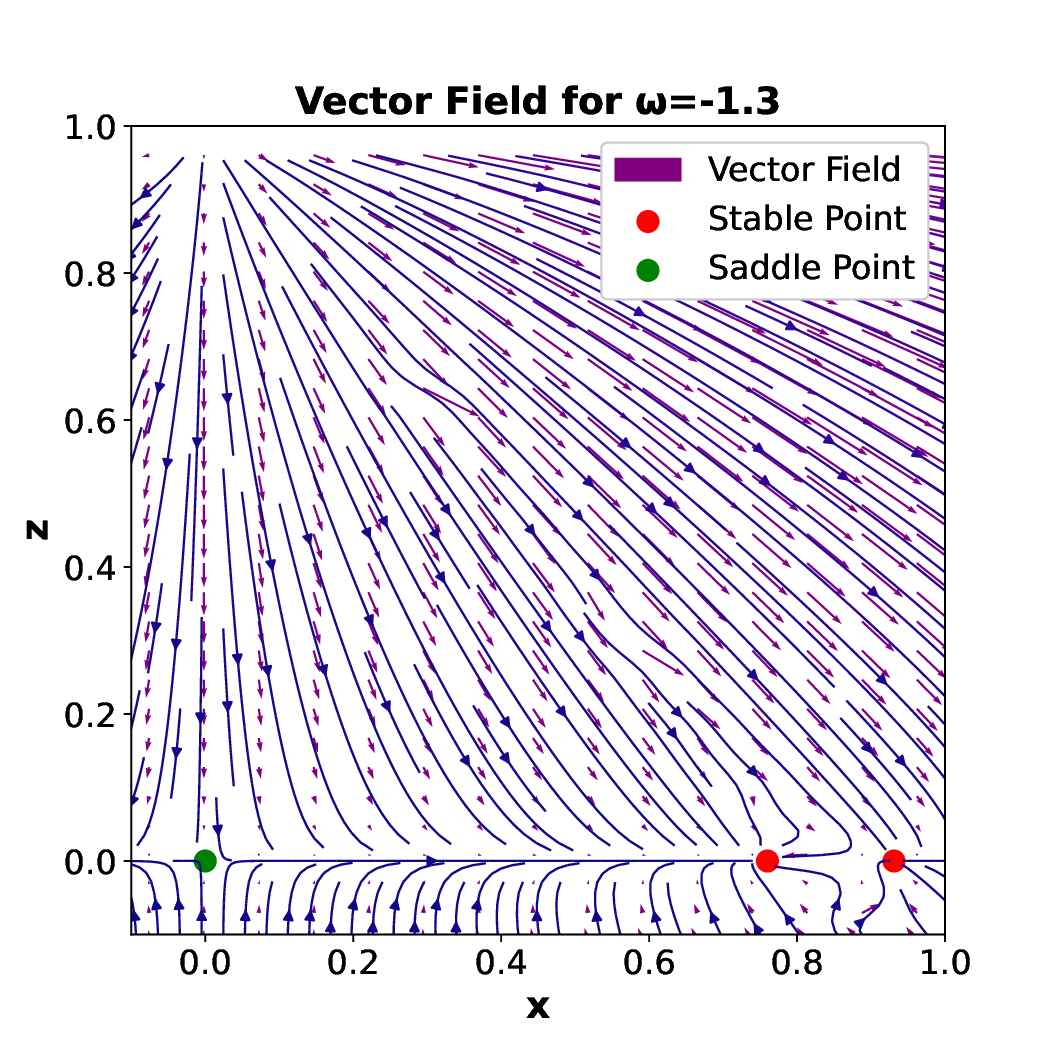}}\\
\subfloat[] {\includegraphics[height=3in,width=3in]{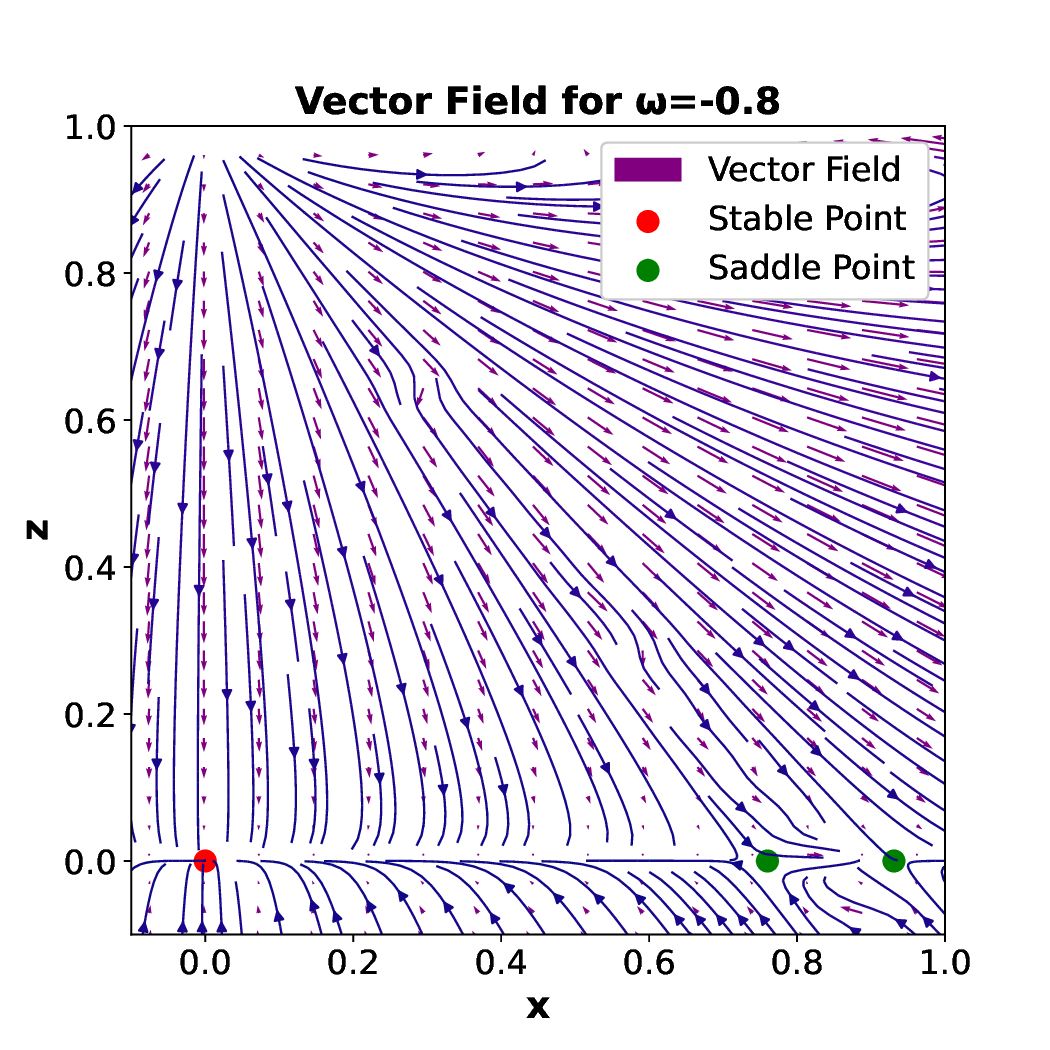}}\hspace{1cm}
\subfloat[] {\includegraphics[height=3in,width=3in]{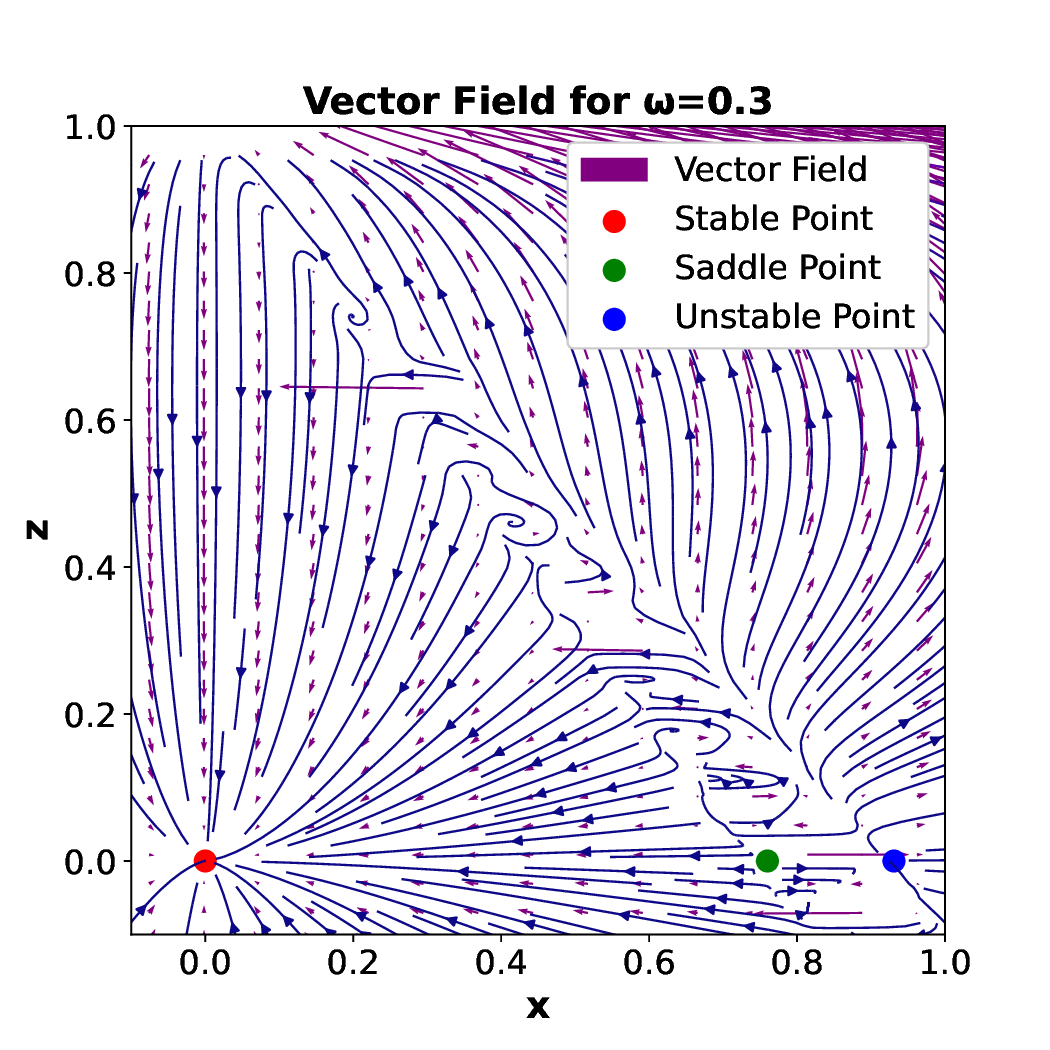}}\\
\caption{Figure $(a)-(d)$ are $z$ vs $x$ phase diagrams for the model $f(Q)=\alpha Q^2+ v Q^2 \log (Q)$ at different values of $\omega$. A red dot indicates a stable point, a green dot indicates a saddle point and a blue dot indicates an unstable point.}
\end{center}
\end{figure} 
We enlist all the fixed points and their categorizations in table 3. Figure 3(a)-3(d) are plots of $x-z$ plane phase portraits for different $\omega$ values. At the same time, two stable nodes and two unstable nodes are observed to form. For one stable point, the dynamical system admits a fixed point where the Hubble parameter H becomes constant or anisotropy turns to vanish or matter density decays. Two distinct stable points can exist when multiple zeros of the effective equation governing $\dot{H}$ are there. Multiple physically allowed solutions for $H$ or $Q$, i.e., multiple de Sitter like attractor may appear. The term $Q^2logQ$ introduces turning point and the equation $2f_Q+Qf_{QQ}=0$ may have two real positive roots for $Q$. On the other hand, the model may include a non-trivial matter sector or anisotropic shear and then one attractor could correspond to matter scaling solution, i.e., deceleration with small anisotropy. Another stable point could be a pure de Sitter attractor with isotropic expansion\cite{ Bahamonde2023, Leon2018_HigherOrderTeleparallel}. Two unstable points may rise due to multiple early time solutions. One of them may correspond to a high $Q$ regime (large Hubble rate), i.e., early inflation. Another may lie at moderate $Q$, representing a matter like or stiff fluid like phase. In presence of double unstable points, the universe is followed to evolve following multiple paths (bifurcation structure) which is similar as the authors of \cite{Prasia2016} have followed.

In the context of the center manifold, take $z$ as the stable variable and $x$ as the central variable (since the eigenvalues are $0$ and $-6$). Let the center manifold theory has the form $z=h(x)$; then the approximation $N$ as,

\begin{equation}
    \begin{split}
        N(h(x))= h^\prime(x) \cdot x^\prime - z^\prime = h^\prime (x) \cdot 3 x \left[\frac{2 x (\omega +1) \left\{16 x+13 (h(x)-1)\right\}}{(h(x)-1) \left\{40 x+33 (h(x)-1)\right\}}+2 \{x \omega +x+h(x)\}-\omega -1\right]\\- [6 h(x) \{x \omega +x+h(x)-1\}]~~~~.
    \end{split}
\end{equation}

For zeroth approximation, this component turns to
$$N(h(x)) = 0 + \mathcal{O} (x^2)~~~.$$

Linearisation around this point yields one vanishing eigenvalue and two negative eigenvalues, indicating that the fixed point is non-hyperbolic, with the zero eigenvalue associated with the shear variable $z$. To analyse the dynamics beyond linear order, we perform a minimal centre manifold reduction.

Introducing perturbations
\begin{equation}
u=z,\qquad v_1=x,\qquad v_2=y-1,
\end{equation}
the system can be written in centre stable form as
\begin{align}
\dot u &= F(u,v),\\
\dot v &= A v + G(u,v),
\end{align}
where $u$ denotes the centre variable, $v=(v_1,v_2)$ represents the stable directions, and the eigenvalues of $A$ have negative real parts.

The centre manifold is locally defined by $v=h(u)$ with $h(0)=h'(0)=0$, implying $h(u)=\mathcal{O}(u^{2})$. Substituting this relation into the evolution equation for $u$ and expanding near $u=0$, the reduced dynamics on the centre manifold takes the form
\begin{equation}
\dot u = \gamma\,\alpha\,u^{2} + \mathcal{O}(u^{3}),
\end{equation}
where $\gamma$ is a nonzero constant. This demonstrates that the evolution along the shear direction is slow and governed by nonlinear terms. The stability of the fixed point depends on the sign of $\alpha$: for $\alpha<0$ the shear decays and the solution isotropises, while for $\alpha>0$ it grows, leading to instability. Small variations of $\alpha$ therefore induce qualitative changes in the phase-space behaviour, supporting the interpretation of bifurcation like features.

Time-time perturbation for this model is given by,
\begin{equation}
    6H\{(2\alpha+v)Q+2vQ logQ\}(\dot{\Psi}+H\Phi)+\{(2\alpha+v)+2vlogQ+2v\}\delta Q = \delta \rho ~~~\text{and}
\end{equation}
space-space perturbation is given by,
\begin{equation}
    2\{(2\alpha+v)Q+2vQ log Q\}(\dot{\Psi}+H\Phi)+\{(2\alpha + v)+2vlog Q + 2v\}(\dot{Q}\Psi+Q\dot{\Psi})= -a^2\delta p ~~~,
\end{equation}
where $\delta Q$ is the non-metricity perturbation given by, 
\begin{equation}
    \delta Q = 12 H(\dot{\Psi}+H\Phi)~~~~.
\end{equation}
The propagation speed of scalar perturbation is,
\begin{equation}
    C_s^2 = \frac{2HQ(v+2\alpha)+\dot{Q}(3v+2\alpha)+2(2HQ+\dot{Q})vlog Q}{6H^3\{(6+Q)v+2(2+Q)\alpha+2(2+Q)vlog Q}~~~,
\end{equation}
stability occurs when $C_s^2>0$ and subliminal when $C_s^2\leq 1$. \\
No-ghost condition is given by, 
\begin{equation}
    (2\alpha+v)Q+2vQlog Q > 0~~ .
\end{equation}
The speed of gravitational waves is,
\begin{equation}
    C_T^2= \frac{v+2\alpha+2vlog Q}{7v+6\alpha+6vlogQ}  ~~~.
\end{equation}
For consistency with (i.e; $C_T^2=1$), we require, 
\begin{equation}
    2Qf_{QQ}<<f_Q  ~~~.
\end{equation}

Under considering the anisotropy and Bianchi-I cosmolgy, the shear ODE is, 
$$\dot{\sigma}_{ij}+\left(3H+\frac{\dot{f}_{Q}}{f_{Q}}\right)\sigma_{ij}=0~~~,$$
i.e., $$\dot{\sigma}_{ij}+\left\lbrace3H+\frac{\dot{Q}}{Q}\frac{2\alpha+v(2lnQ+3)}{2\alpha+v(2lnQ+1)}\right\rbrace\sigma_{ij}=0~~~.$$

This gives the solution,
$$\sigma_{ij}(t)=\sigma_{ij}(t_{*})a^{-3}(t)a^3(t_{*})\frac{f_{Q}(t_{*})}{f_{Q}(t)}~~~.$$

Thus,
$$\left(\frac{\sigma_{*}}{H}\right)_{0}=\frac{\sigma_{*}}{H_{0}}a^{-3}(t_{0})\frac{f_{Q}(t)}{f_{Q}(t_{0})}\lesssim 5\times10^{-11} ~~~\text{(for $95\%$ confidence-level (CL))}~~~, $$

where the reference time $t_{*}$ can be taken as a convenient normalization time (e.g; recombination $z\sim1100$ or earlier).\cite{saadeh2016isotropic}

%%%%%%%%%%%%%%%%%%%%%%%%%%%%%%%%%%%%%%%%%%%%%%%%%%%%%%%%%%%%%%%%%%%%%%

\subsection{Model-IV : $f(Q)=\eta Q_0 \sqrt{\frac{Q}{Q_0}} \log\left(\frac{\lambda  Q_0}{Q}\right)$}
The fractional power $Q^{\frac{1}{2}}$ dominates at low curvature at late-times. A logarithmic correction $log\left(\frac{Q_0}{Q}\right)$ is reminiscent of quantum loop or renormalisation corrections \cite{Nojiri2004_lnR}. A prefactor $\eta Q_0$ sets a cosmological scale. Such structure arises in non-local or scale dependent gravity, renormalisation group which improves effective actions. This model with conformal anomalies or vacuum polarisation. In Bianchi-I, where shear evolves as $\dot{\sigma}+3H\sigma$, the slowly decaying logarithmic correction modifies the friction term for shear\cite{barrow2006anisotropically}.

Here, $\eta$ is a dimensionless constant, $Q_0$ is a constant scale of non-metricity, $\lambda$ is a dimensionless parameter and $Q$ is the non-metricity scalar. The $ Q$-derivatives of the model are,
\begin{equation}\label{Model IV f_Q}
    f_Q = {\frac{\eta}{2} \sqrt{\frac{Q_0}{Q}} \left\lbrace{\log \left(\frac{\lambda  Q_0}{Q}\right)-2}\right\rbrace}~~~\text{and}
\end{equation}
\begin{equation}\label{Model IV f_QQ}
    f_{QQ} = -{\frac{1}{4} \frac{\eta}{Q}\sqrt{\frac{Q_0}{Q}}  \log \left(\frac{\lambda  Q_0}{Q}\right)}~~~.
\end{equation}
Substituting the value of $f_Q$ in equation \eqref{expression of Q}, we obtain,
\begin{equation}
    \log \left(\frac{\lambda Q_0}{Q}\right) = \frac{2 (1-z)}{4x+3z-3}~~.
\end{equation}
Substituting the values of equations \eqref{Model IV f_Q} and \eqref{Model IV f_QQ}, we can easily find the value of $\Gamma$ as,
\begin{equation}
    \Gamma = \frac{8(x+z-1)}{1-z}~~.
\end{equation}
Now the dynamical variables are reduced to,
\begin{equation}
    x^\prime = 3 x \left\{\frac{x (\omega +1)}{3-4 x-3 z}+2 (x \omega +x+z)-\omega -1\right\}~~~,
\end{equation}
\begin{equation}
    z^\prime =6 z (x \omega +x+z-1)~~~\text{and}
\end{equation}
equation \eqref{Reduced H equation} reduced to,
\begin{equation}
    \frac{\dot{H}}{H^2} = -\frac{3 \left(4 x^2 (\omega +1)-4 x (\omega +1)+x (3 \omega +7) z+3 (z-1) z\right)}{4 x+3 z-3}~~~.
\end{equation}

For this model, the exponential correction saturates at late times, and the system approaches a stable accelerating fixed point with $\dot H \rightarrow 0$. As a result,
\begin{equation}
\omega_{\mathrm{eff}} \rightarrow -1 ,
\end{equation}
corresponding to a robust dark energy dominated attractor. During intermediate stages, $|\dot H|\ll H^{2}$, so that $\omega_{\mathrm{eff}}$ remains close to $-1$, describing a smooth transition toward accelerated expansion.

See Appendix-II (B.4) for a brief summary of derived expressions.
\begin{table}[h!]
    \centering
    \begin{tabular}{|>{\centering\arraybackslash}m{1.5cm}|>{\centering\arraybackslash}m{1.5cm}|>{\centering\arraybackslash}m{2.2cm}|>{\centering\arraybackslash}m{2.4cm}|>{\centering\arraybackslash}m{3.1cm}|>{\centering\arraybackslash}m{1.6cm}|>{\centering\arraybackslash}m{2.4cm}|}
    \hline
       {\bf Crit.} & {\bf Exist-} & {\bf Eigenvalues} & {\bf Type of crit.} & {\bf Stability} & {\bf $C^2_{T}$} & {\bf Cosmology}\\
       {\bf Point} & {\bf ence} & & {\bf point} & & &\\
       \hline
       $x\to0, z\to 0$ & $\forall \omega$ & $-3 (\omega+1)$ & Hyperbolic & Saddle for & & $a(t)=a_0 e^{H_0 t}$\\
       && and $-6$ & $(\omega \neq -1)$ & $\omega<-1$ & & (i.e., de Sitter)\\
       \cline{5-5}
       & & & & Stable Node for & & \\
       &&&& $\omega >-1$ & &\\
       \cline{4-5}
       & & & Non-hyperbolic & Center manifold & & \\
       &&& $(\omega=-1)$ & for $\omega =-1$ & & \\
       \cline{1-5}\cline{7-7}
       $x\to\frac{3}{8}, z\to 0$ & $\forall \omega$ & $\frac{3}{4} (3 \omega -5)$ and  & Hyperbolic &  Stable Node & & $a \propto t^{\frac{8}{15(1+\omega)}}$\\
       && $\frac{15}{4} (\omega+1)$ & $(\omega \neq -1, \frac{5}{3})$ & for $\omega<-1$ & &\\
       \cline{5-5}
       & & & & Saddle point for & & \\
       &&&& $-1<\omega <\frac{5}{3}$ & & \\
       \cline{5-5}
       &&&& Unstable Node & &\\
       &&&& for $\omega>\frac{5}{3}$ & &\\
       \cline{4-5}
       &&& Non-hyperbolic & Unstable for & $\frac{log \left(\frac{\lambda Q_0}{Q}\right)-2}{4}$ &\\
       &&& $(\omega = -1, \frac{5}{3})$ & $\omega = \frac{5}{3}$ & &\\
       \cline{5-5}
       & & & & Center manifold & & \\
       & & & & for $\omega=-1$ & & \\
       \cline{1-5}\cline{7-7}
       $x\to1, z\to 0$ & $\forall \omega$ & $6 \omega$ and & Hyperbolic & Stable node for & & $a(t)=a_0 e^{H_0 t}$\\
       && $15 (\omega +1)$ & $(\omega \neq -1, 0)$ & $\omega<-1$ & & (i.e., de Sitter)\\
       \cline{5-5}
       &&&& Saddle point for & &\\
       &&&& $-1<\omega<0$ & &\\
       \cline{5-5}
       &&&& Unstable node & &\\
       &&&& for $\omega>0$ & &\\
       \cline{4-5}
       &&& Non-hyperbolic & Unstable for & &\\
       &&& $(\omega =-1, 0)$ & $\omega =0$ & &\\
       \cline{5-5}
       &&&& Center manifold && \\
       &&&& for $\omega=-1$ && \\
       \hline
    \end{tabular}
    \caption{Critical points with existence conditions, eigenvalues, type of critical points, stability, cosmology pattern and value of $C^2_{T}$ for model-IV : $f(Q)=\eta Q_0 \sqrt{\frac{Q}{Q_0}} \log\left(\frac{\lambda  Q_0}{Q}\right)$ ( see Appendix-I (A.4) for detailed calculation).}
\end{table}

\begin{figure}[h!]
\begin{center}
\subfloat[] {\includegraphics[height=3in,width=3in]{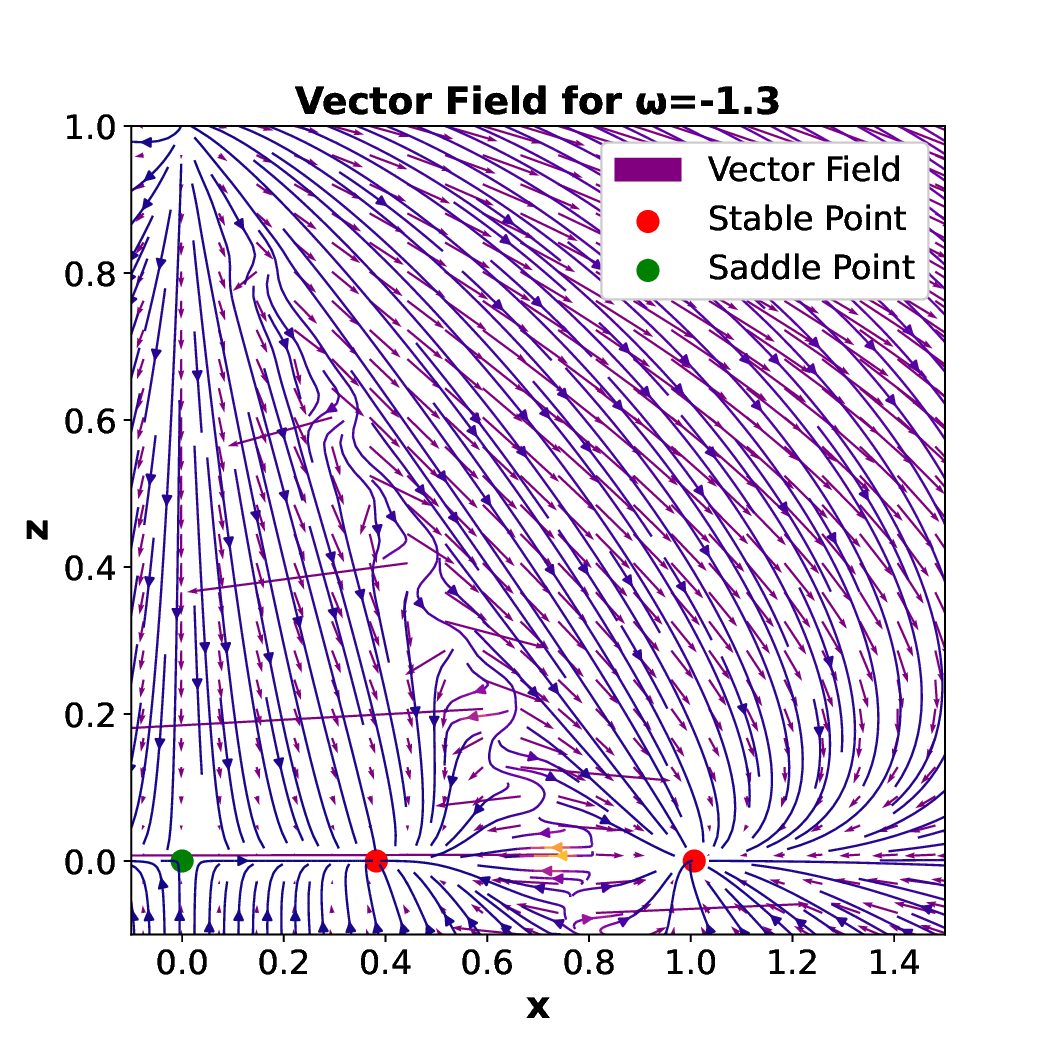}}\hspace{1cm}
\subfloat[] {\includegraphics[height=3in,width=3in]{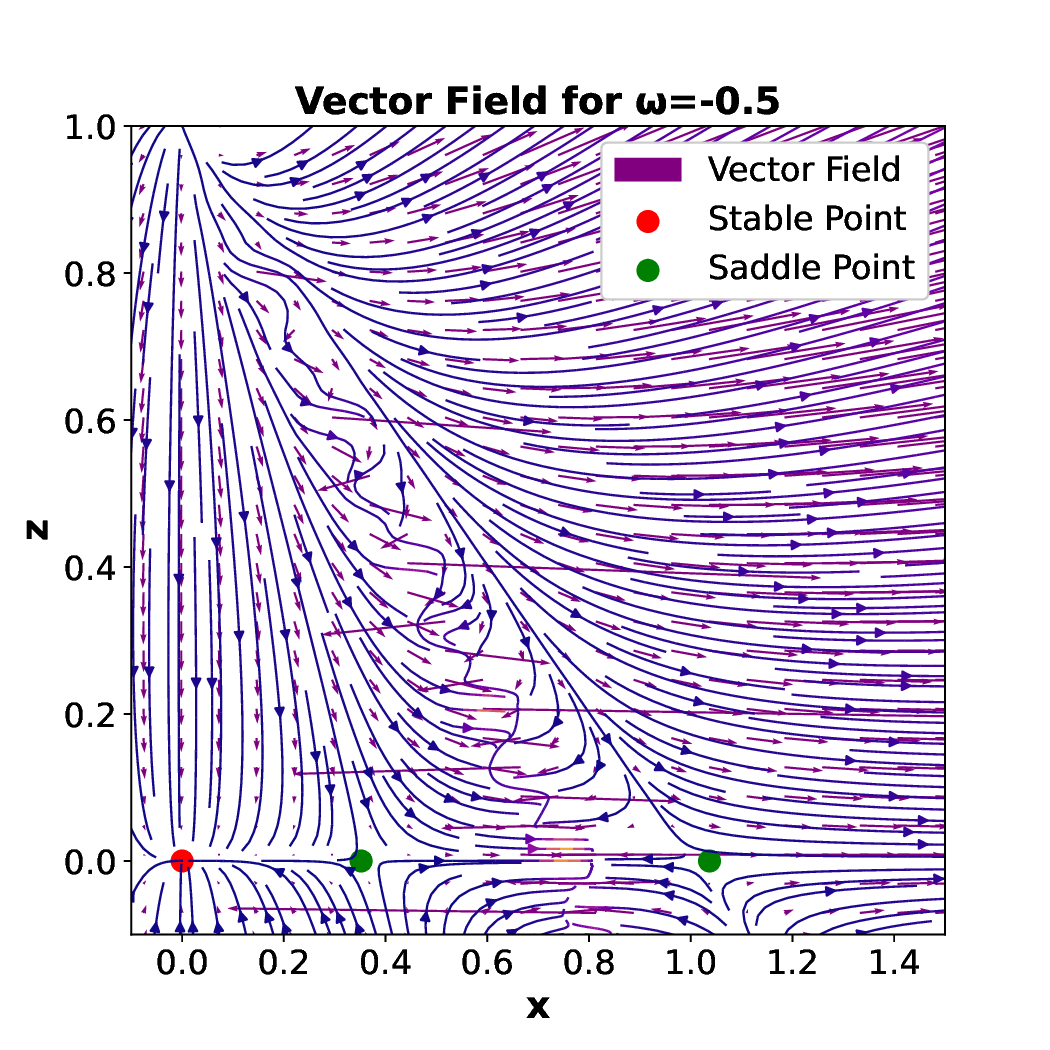}}\\
\subfloat[] {\includegraphics[height=3in,width=3in]{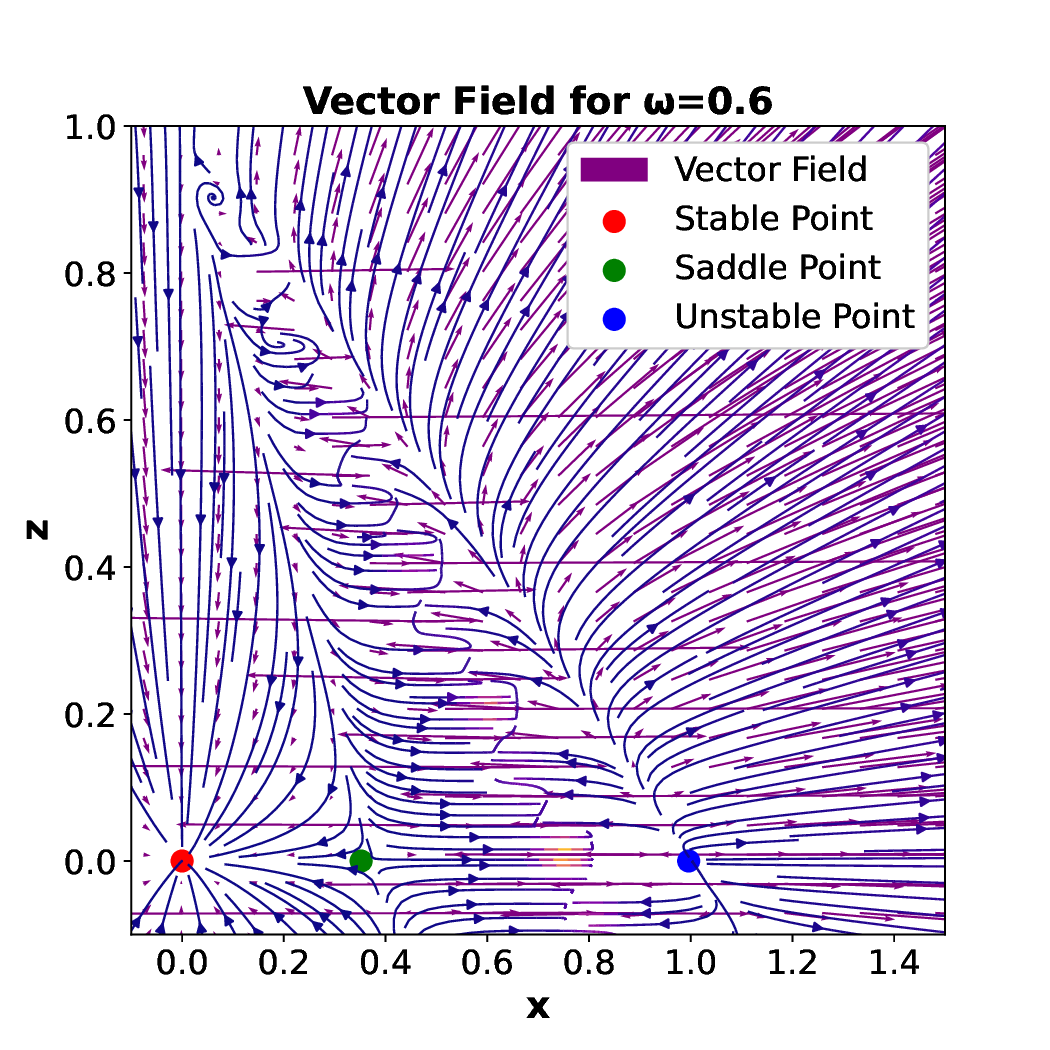}}\hspace{1cm}
\subfloat[] {\includegraphics[height=3in,width=3in]{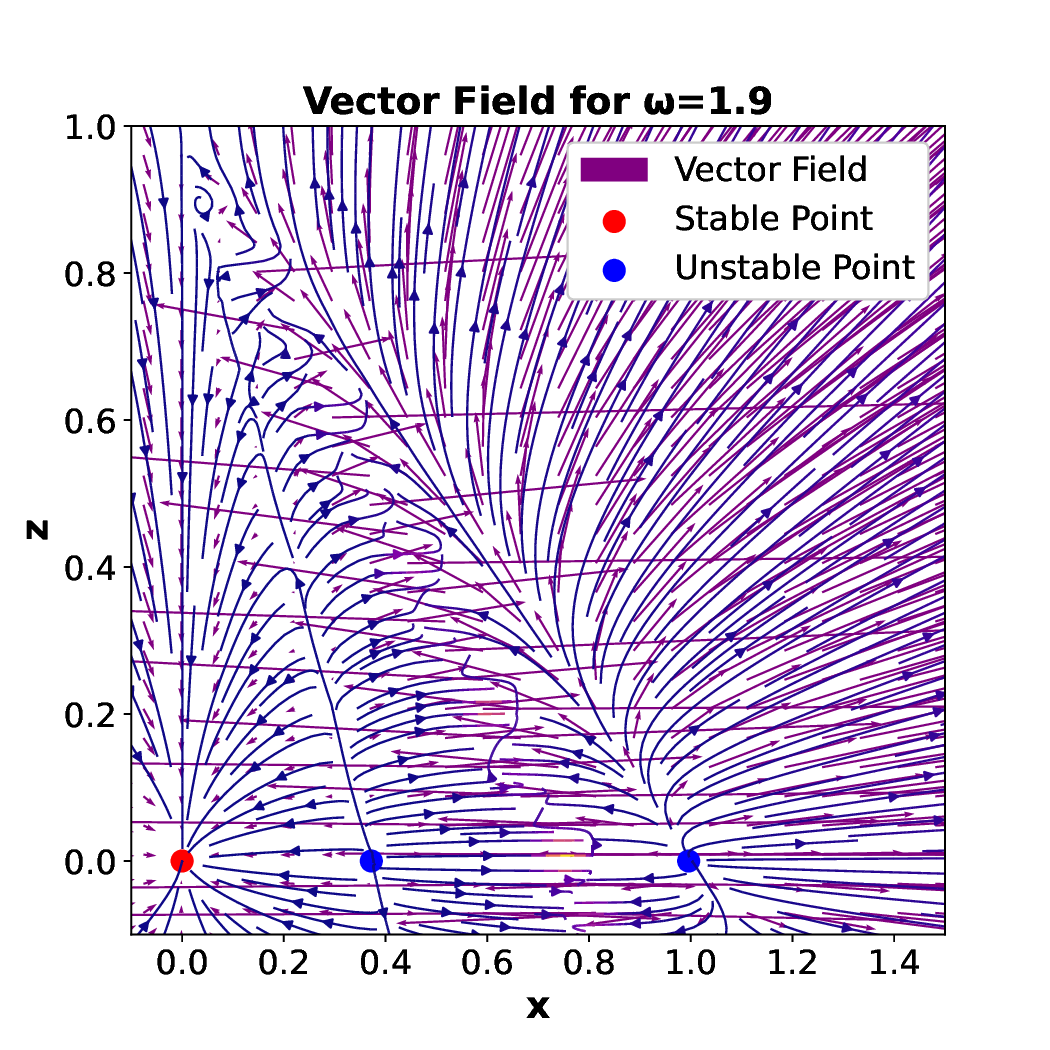}}\\
\caption{Figure $(a)-(d)$ are $z$ vs $x$ phase diagrams for the model $f(Q)=\eta Q_0 \sqrt{\frac{Q}{Q_0}} \log\left(\frac{\lambda  Q_0}{Q}\right)$ at different values of $\omega$. A red dot indicates a stable point, a green dot indicates a saddle point and a blue dot indicates an unstable point.}
\end{center}
\end{figure}
In table 4, different critical points and their categorizations are noted. Fig $4a - 4d$ are $z~\text{vs}~x$ phase diagrams for different values of $\omega$. At the same time, stable, unstable and saddle nodes are found to form. Multiple de Sitter like attractors, i.e., physically acceptable solutions appear. de Sitter solutions appear at $x \to 0$ and $x\to1$ for different critical point, suggesting that model IV supports the accelerating expansion as stable or unstable phases. Late-time cosmic acceleration as a de Sitter attractor is found at $x \to 0$ and the possible inflationary era as a de Sitter solution is formed at $x \to 1$. A $x\to\frac{3}{8}$ power-law expansion appears, which may correspond to matter/radiation dominated era depending on the values of $\omega$. The stability of each phase depends strongly on the EoS $\omega$, i.e., this model can describe different cosmological epochs dynamically. The non-hyperbolic points correspond to the bifurcation or marginal stability. So these points required the center manifold or higher order analysis. 

To calculate the center manifold for this model take $z$ as the stable variable and $x$ as the central variable. Let the center manifold theory has the form $y=h_1(x)$ and $u=h_2(x)$; then the approximation $N$ has one component:

\begin{equation}
\begin{split}
N(h(x))= h^\prime(x) \cdot x^\prime - z^\prime = h^\prime (x) \cdot 3 x \left[\frac{x (\omega +1)}{3-4 x-3 h(x)}+2 \{x \omega +x+h(x)\}-\omega -1\right] \\-[6 h(x) \{x \omega +x+h(x)-1\}]~~~~.
\end{split}
\end{equation}

For the zeroth approximation,
$$N(h(x)) = 0+ \mathcal{O}(x^2) ~~~. $$

Introducing perturbations
\begin{equation}
u=z,\qquad v_1=x,\qquad v_2=y-1,
\end{equation}
the system can be written in centre stable form as
\begin{align}
\dot u &= F(u,v),\\
\dot v &= A v + G(u,v),
\end{align}
where $u$ denotes the centre variable and the eigenvalues of $A$ have negative real parts.

The centre manifold is locally defined by $v=h(u)$ with $h(0)=h'(0)=0$, implying $h(u)=\mathcal{O}(u^{2})$. Substituting this relation into the evolution equation for $u$ and expanding near $u=0$, the reduced dynamics on the centre manifold takes the form
\begin{equation}
\dot u = -\gamma\,e^{-Q_\ast/Q_{0}}\,u^{2} + \mathcal{O}(u^{3}),
\end{equation}
where $\gamma>0$ is a constant. This demonstrates that the evolution along the shear direction is slow and asymptotically stable. The exponential suppression ensures decay of anisotropy and confirms the robustness of the late time de Sitter attractor in Model IV.

Like model III, model IV is also observed to possess two stable or two unstable fixed nodes at a time. They are followed to lie on the $z=0$ axis. 

The time-time perturbation emerges as,
\begin{equation}
    3Hn \sqrt{\frac{Q_0}{Q}}\left\lbrace{log \frac{\lambda Q_0}{Q}-2}\right\rbrace\left\lbrace\dot\Psi + H\Phi\right\rbrace -\frac{n}{4}Q_{0}^{\frac{1}{2}}Q^{-\frac{3}{2}}ln\left(\frac{\lambda Q_{0}}{Q}\right)\delta Q = \delta \rho~~~.
\end{equation}
Space-space perturbation is given by,
\begin{equation}
    n\sqrt{\frac{Q_0}{Q}}\left\lbrace{log \left(\frac{\lambda Q_0}{Q}\right)-2}\right\rbrace \left(\dot\Psi+H\Phi \right)-\frac{n\sqrt Q_0}{4Q^\frac{3}{2}}log \left(\frac{\lambda Q_0}{Q} \right)\times \left(\dot Q \Psi + Q \dot \Psi \right) = -a^2 \delta p ~~~,
\end{equation}
where $\delta Q$ is the non-metricity perturbation given by $\delta Q = 12 H(\dot \Psi + H \Phi)$ .
The propagation speed of scalar perturbation is, 
\begin{equation}
    C_s^2 = \frac{8HQ+\left(\dot Q -4HQ \right)log \left(\frac{\lambda Q_0}{Q} \right)}{12H^2 \left\lbrace{2Q+(H-Q)log \left(\frac{\lambda Q_0}{Q}\right)}\right\rbrace}~~~.
\end{equation}
For stability condition $0 \leq C_s^2 \leq 1$, no-ghost condition is $\frac{n}{2}\sqrt{\frac{Q_0}{Q}}\left\lbrace{ log \left(\frac{\lambda Q_0}{Q}\right)-2}\right\rbrace>0$ and the speed of gravitational waves is, 
\begin{equation}
    C_T^2 = \frac{log \left(\frac{\lambda Q_0}{Q}\right)-2}{4}~~~~.
\end{equation}
For $C_T^2 \approx1 $ (as in GR), we required, $log \left(\frac{\lambda Q_0}{Q} \right) \approx 2~\Rightarrow Q\approx \lambda Q_0e^2$, which may conflict with the no-ghost condition unless $\lambda$ is fine-tuned.

The shear evolution equation for this model is, 
\begin{equation*}
    \dot{\sigma}_{ij} + \left[{3H + \frac{\dot{Q}}{Q} \frac{{-\frac{1}{4} ln \left({\frac{\lambda Q_0}{Q}}\right)}}{{\frac{1}{2} ln\left({\frac{\lambda Q_0}{Q}}\right) -1}}}\right] \sigma_{ij} = 0~~~,
\end{equation*}
considering the anisotropy and Bianchi-I.

At a late-time, $Q<< \lambda Q_0$, then $ln \left({\frac{\lambda Q_0}{Q}}\right)>>1$. Then, 
\begin{equation*}
    \frac{\dot{f}_Q}{f_Q} \approx -\frac{1}{2} \frac{\dot{Q}}{Q} =-\frac{3}{2} H ~~~~\left(\text{as}~\frac{\dot{Q}}{Q}=3H\right)~~~.
\end{equation*}
Then, $\dot{\sigma}_{ij} + (3H +1.5H)\sigma_{ij}=0~~~\implies \sigma \propto a^{-4.5}$.

Then between recombination $(a_{*} \simeq 10^{-3})$ and today,
\begin{equation*}
    \frac{\sigma_0}{\sigma_*} \sim (10^{-3})^{4.5} = 10^{-13.5}~~~,
\end{equation*}
where $a_*$ is the scale factor at the reference time $t_*$, $\sigma_0$ is the shear scalar at time $t_0$ and $\sigma_*$ is the shear scalar at time $t_*$.

So, even if $\left({\frac{\sigma}{H}}\right)_* \sim 10^{-2}$ at recombination then, 
\begin{equation*}
    \left({\frac{\sigma}{H}}\right)_0 \sim 10^{-15.5} << 5\times 10^{-11}~~~\text{[i.e., within observation limit]}~~~,
\end{equation*}
where $()_*$ denotes the value at the reference time $t_*$

Hence, this $f(Q)$ produces super-efficient isotropization at late-times.\cite{saadeh2016isotropic}

%%%%%%%%%%%%%%%%%%%%%%%%%%%%%%%%%%%%%%%%%%%%%%%%%%%%%%%%%%%%%%%%%%%
\section{Brief Discussion and Conclusion}
%%%%%%%%%%%%%%%%%%%%%%%%%%%%%%%%%%%%%%%%%%%%%%%%%%%%%%%%%%%%%%%%%%%
Early universe measurements of the value of Hubble parameter differs from the same measured for late-time universe. Hubble tension measured thus suggests regarding the incompleteness of standard general relativity or cosmological principle. Different voids, temperature fluctuations, dipole modulation, anisotropic inflation etc. points towards anisotropic universe. A curvature-free torsion-free gravity is chosen. The signature entity will be the non-metricity scalar $Q$. It is followed that this gravity theory, depending on different free parameters of the model, can fit early inflation to late-time cosmic acceleration. No need to modify the stress energy tensor part of field equation with any exotic fluid is required. Our motive was to analysis different fixed points of the phase portraits constructed upon this model. For this, some model parameters are carefully chosen so that their derivatives are found to linearly depend on the same set of parameters. Vanishing derivatives indicates extremality and hence we check for singular properties of the corresponding Jacobian matrix is followed to point properties and classifications of the fixed/ critical points. Cosmological implications are investigated. 

First gravity model chosen with non-metricity scalar is $f(Q)=mQ^n$. This mimics general relativity for $m=n=1$. $n>1$ is able to reproduce early inflation. For $0<n<1$, however, gravity gets weaken and can be interpreted as the effect of a repulsive force, i.e., the late-time cosmic acceleration is made after. This model can predict gravitational waves to propagate with the speed of light which is supported by GW170817 event \cite{abbott2017gw170817, 2017ApJabott}. We have enlisted the fixed points for this model in table 1 and phase portraits are emphasised in fig 1a-1d. We find hyperbolic stable, unstable nodes and saddles. Non-hyperbolic critical point, for this model comes as a center manifold. Among these, the hyperbolic stable point, formed near the origin in all phase portraits signifies that the universe should always end up in the same state. This also explains why today's universe is isotropic to a large scale and accelerating. Also stable point corresponding to $x=0$ indicates constant Hubble parameter and hence a late-time acceleration.

In Bianchi-I, a saddle point having non-zero anisotropy can start from a anisotropic state and exit towards an isotropic attractor. Hyperbolic unstable points may indicate that the universe starts near this kind of points(eg, anisotropic inflation), strong repulsion may cause nearby trajectories to diverge rapidly and create an unstable fixed point. Shear variable from anisotropic contribution can lead so. Lastly, we must remember, an unstable node not only be thought to express late-time attractor. It may also signify a past attractor as well.

In the second model, $f(Q)=e^{nQ}$ in Bianchi-I cosmology provides a minimal, purely geometric mechanism for explaining the universe's anisotropic origin and accelerated fate, all within a unified, covariant and second order field theory. $f(Q)=e^{nQ}$ grows rapidly, so does $f_Q$, leading to strong geometric feedback on the shear evolution. If matter is subdominant and $n>0$, shear decays but slowly which implies anisotropic inflation. Linear perturbations grow away from this node such that the system exists anisotropic phase naturally. Besides such unstable points, if we look towards stable nodes, we follow the shear equation $\dot{\sigma}+3h\sigma \approx 0$ and to imply $\sigma \propto e^{-3ht}$ and the universe ends in a fully isotropic accelerated expansion, consistent with CMB constraints. At late-time shear vanishes to restore isotropy.

In this model, fully degenerate fixed point is found. Hartman Grobman theorem does not apply for these points. Trajectories near this point behave non-linearly. Universe near such points may exhibit slow evolution, neither accelerating nor decelerating strongly. Small anisotropy and matter or quantum fluctuations may push towards different attractors. This may again represent bifurcation thresholds, eg., transition from isotropic to anisotropic solutions or inflation to matter era. 

A center manifold is also observed to form. This lies in the shear direction while other parameters are stable. Here shear neither grows nor decays first. Universe remains mildly anisotropic. Higher order corrections regulate whether the universe becomes isotropic later or not. Universe could hover near the manifold, mimicking near-isotropic expansion.

The third model considered is $f(Q)=\alpha Q^2+v Q^2logQ$. If $Q$ is small and positively vanishes, $Q^2logQ$ term also does so and $Q^2$ dominates representing a de Sitter attractor/phantom. At high $Q$, inflation like dynamics is expected. While finding and categorizing fixed points, both hyperbolic and non-hyperbolic fixed points are seen to be formed. More than one stable points are followed. One of them is a stable node which signifies early inflation like quasi de Sitter phase. Another is a stable spiral pointing towards late-time DE dominated de Sitter space. Geometric tracking solutions are interpreted by centers. In a phase portrait, we have followed two unstable nodes. One may signify the early universe saddle for high shear and the rest one may be caused by inflationary node. The actual trajectory should depend on the initial shear, matter content and the sign of different model parameters.

Last studied model is $f(Q)=\eta Q_0 \sqrt{\frac{Q}{Q_0}} \log\left(\frac{\lambda  Q_0}{Q}\right)$. In the Bianchi-I universe, chosen model provides a compelling geometric framework for exploring anisotropic cosmology, combining the effects of square-root and logarithmic non-metricity terms. Physically, it allows the universe to transit smoothly from an early anisotropic matter dominated phase to a late-time accelerated and isotropic expansion without invoking dark energy fields. The logarithmic correction enhances the model's flexibility at low $Q$ (late-times), enabling de Sitter like behavior, while the square root term ensures controlled dynamics at high $Q$ (early times). In the Bianchi-I background, it governs the decay of anisotropy, potentially hosting stable attractors and metastable center manifolds that regulate cosmic evolution. Such a model is valuable for explaining cosmic acceleration, isotropization and transitions between epochs entirely through gravitational geometry, making it a strong candidate for a unified description of the universe’s expansion history. Overall, the dynamical model matches the same for the third model to a large extent.

Although a full confrontation with observational data is not attempted here, the viability conditions $c_T^2 \simeq 1$ and $c_s^2 > 0$ impose non-trivial restrictions on the model parameters. These constraints already exclude certain branches of solutions and significantly reduce the allowed parameter space, highlighting the phenomenological relevance of the stability analysis.

The present study should therefore be viewed as a bridge between generic dynamical system formulations of $f(Q)$ gravity and more detailed phenomenological or observational analyses. By systematically comparing several representative models in an anisotropic setting, we identify common qualitative features and potential pitfalls that can guide future, more refined investigations.

We therefore stress that the perturbative analysis presented here is exploratory in nature and intended to provide indicative consistency checks rather than a definitive assessment of gravitational-wave propagation in anisotropic $f(Q)$ cosmologies.

Before concluding, it is useful to summarise which regions of parameter space remain theoretically viable once the various consistency requirements discussed in this work are taken into account. These include the absence of ghost instabilities, positivity of the scalar sound speed $c_s^2$, acceptable tensor propagation speed $c_T^2 \simeq 1$, and the existence of a stable late-time accelerating attractor with controlled anisotropy.

For model I, theoretical viability strongly favours parameter values close to the general relativistic limit $n\simeq 1$. Significant deviations from $n=1$ tend to be disfavoured by either ghost or gradient instabilities, or by the absence of a stable late-time accelerating attractor. While small departures from $n=1$ may be relevant for early-time or intermediate dynamics, the late-time viability region is narrow.

In Model II, viable regions of parameter space correspond to values of $m$ for which the scalar sector remains ghost-free and $c_s^2 \ge 0$, while admitting a late-time accelerating fixed point. Large values of $|m-1|$ are typically disfavoured, as they either destabilise the scalar perturbations or lead to growing anisotropies.

For the logarithmic model, the viability conditions restrict the sign and magnitude of the logarithmic coupling parameter. Only a limited region of parameter space satisfies the ghost-free condition, positive sound speed, and a stable accelerating attractor. Outside this region, the models exhibit either unstable perturbations or pathological late-time behaviour.

Among the models considered, the exponential form exhibits the most robust theoretical behaviour. A broad region of parameter space satisfies all the consistency requirements, including positive sound speed, absence of ghosts, acceptable tensor speed, and a stable de Sitter like attractor with decaying anisotropy. This makes the exponential model particularly appealing from the viewpoint of theoretical viability.

Overall, the combined stability and perturbative analyses indicate that only restricted regions of parameter space are theoretically viable for most models, with the exponential model admitting the widest allowed region. These results should be viewed as necessary theoretical consistency conditions, which any further phenomenological or observational study must satisfy. A brief summary table is also performed in Appendix-III.

%%%%%%%%%%%%%%%%%%%%%%%%%%%%%%%%%%%%%%%%%%%%%%%%%%%%%%%%%%%%%%%%%%%%%
\section*{CRediT authorship contribution statement}

{\bf Subhajit Pal:} Writing -- review ~\& ~editing, Writing -- original draft, Visualization, Software, Methodology, Investigation, Formal analysis, Conceptualization. {\bf Atanu Mukherjee:} Visualization, Investigation, Conceptualization. {\bf Ritabrata Biswas:} Writing -- review ~\& ~editing, Writing -- original draft, Visualization, Validation, Methodology, Investigation, Formal analysis, Conceptualization. {\bf Farook Rahaman:} Editing, Methodology, Conceptualization. 

\section*{Acknowledgment} The authors wish to thank IUCAA, Pune, where the major part of this article was constructed. RB and FR thank IUCAA, Pune, for the Visiting Associateship. 

\section*{Data Availability Statement}
Data sharing not applicable to this article as no datasets were generated or analysed during the current study.

\section*{Conflict of Interest}

There are no conflicts of interest.

\section*{Funding Statement}

There is no funding to report for this article.

\section*{Code/Software}

No software/Coder was used in this study.

\newpage

\appendix

\section{Appendix-I}

Suppose a vector-valued function is given by,
$$f(x_{1},x_{2})=\begin{bmatrix}
    f_{1}(x_{1},x_{2}) \\ f_{2}(x_{1},x_{2})
\end{bmatrix} ~~~~~.$$ 
The Jacobian matrix linearizes the system around an equilibrium point and is defined as (considering 2D system):
    $$J\left(\frac{f_{1},f_{2}}{x_{1},x_{2}}\right)= \left[\frac{\partial f_i}{\partial x_j}\right]=\begin{bmatrix}
\frac{\partial f_1}{\partial x_1} & \frac{\partial f_1}{\partial x_2} \\
\frac{\partial f_2}{\partial x_1} & \frac{\partial f_2}{\partial x_2} 
\end{bmatrix} ~~~~,$$ 
where $i,j = 1,2$ ~~.\\

For the functions $x^\prime$ and $z^\prime$, the Jacobian matrix takes the form
\begin{equation}
    J\left(\frac{x^\prime,z^\prime}{x,z}\right)=\begin{bmatrix}
\frac{\partial x^\prime}{\partial x} & \frac{\partial x^\prime}{\partial z} \\
\frac{\partial z^\prime}{\partial x} & \frac{\partial z^\prime}{\partial z}
\end{bmatrix}~~~.
\end{equation}

\subsection{Model I : $f(Q) = mQ^n$}
\begin{equation}
    x^\prime = 3 x \left\{{2 (x + z + x \omega)+\frac{
   2 x (1 + \omega)}{\left(2 + \frac{1}{n-1}\right) (z-1)} -1 - \omega}\right\}~~~ \text{and}
\end{equation}
\begin{equation}
    z^\prime = 6 z (x \omega +x+z-1)~~~.
\end{equation}
The critical points are $(0,0)$ and $\left(\frac{2n-1}{2n},0\right)$.
To solve the Jacobian matrix, we calculate the following :
\begin{equation}
    \frac{\partial x^\prime}{\partial x} =3x\left\{2(1+\omega)+\frac{2(1+\omega)}{\left(2+\frac{1}{n-1}\right)(z-1)}\right\}+3\left\{2(x+z+x\omega)+\frac{2x(1+\omega)}{\left(2+\frac{1}{n-1}\right)(z-1)}-1-\omega\right\}~~~,
\end{equation}
\begin{equation}
     \frac{\partial x^\prime}{\partial z}=3x\left\{2-\frac{2x(1+\omega)}{\left(2+\frac{1}{n-1}\right)(z-1)^2}\right\}~~~,
\end{equation}
\begin{equation}
    \frac{\partial z^\prime}{\partial x}=6z(1+\omega)~~~~~\text{and}
\end{equation}
\begin{equation}
    \frac{\partial z^\prime}{\partial z}=6z+6(x\omega+x+z-1)~~~~.
\end{equation}
Calculating the Jacobian matrix at the critical point $(0,0)$, we obtain,

\begin{equation}
    J\Bigm\vert_{(0,0)}=\begin{bmatrix}
-3(1+\omega) & 0 \\
0 & -6 
\end{bmatrix}~~~.
\end{equation}
The eigenvalues of the above Jacobian matrix are defined as, 
\begin{equation}
    |J-\lambda I |\Bigm\vert_{(0,0)} = 0 ~~~
    \implies \lambda = -3(1+\omega)~~ \text{and}~~ -6,~~~~~~,~~~~~ \forall \omega~~~.
\end{equation}
Jacobian matrix at the critical point $\left(\frac{2n-1}{2n},0\right)$ turns

\begin{equation}
    J\Bigm\vert_{\left(\frac{2n-1}{2n},0\right)}=\begin{bmatrix}
3(1+\omega) & 6\left(\frac{2n-1}{2n}\right)\left\lbrace{1-\frac{n-1}{2n}(1+\omega)}\right\rbrace \\
0 & \frac{3}{n}\{(2n-1)\omega-1\} 
\end{bmatrix}~~~.
\end{equation}
Eigenvalues are calculated as,
\begin{equation}
        \implies \lambda = 3(1+\omega) ~~\text{and}~~ \frac{3}{n}\{(2n-1)\omega-1\} ~~~,~~~ \forall \omega~~~.
\end{equation}
\subsection{Model-II : $f(Q)= exp \{nQ\}$}

\begin{equation}
    x^\prime = \frac{6 x \left\{x^2 (\omega +1)-2 x (\omega +1)+x (2 \omega +3) z+(z-1) (2 z-\omega -1)\right\}}{x+2 z-2}~~~\text{and}
\end{equation}
\begin{equation}
    z^\prime = 6 z (x \omega +x+z-1)~~~.
\end{equation}

The critical points are $(0,0)$ and $(1,0)$.
Solving the Jacobian matrix, 
\begin{equation}
\begin{split}
    \frac{\partial x^\prime}{\partial x} = \frac{6x[2x(1+\omega)+z(3+2\omega)-2(1+\omega)]}{x+2z-2} &-\frac{6x[(z-1)(2z-1-\omega)+x^2(1+\omega)+xz(3+2\omega)-2x(1+\omega)]}{(x+2z-2)^2} \\
    &+ \frac{6[(z-1)(2z-1-\omega)+x^2(1+\omega)+xz(3+2\omega)-2x(1+\omega)]}{x+2z-2}~~,
\end{split}
\end{equation}
\begin{equation}
    \frac{\partial x^\prime}{\partial z}=\frac{6x[x^2+4(z-1)^2+x(4z-5-\omega)]}{(x+2z-2)^2}~~~,
\end{equation}
\begin{equation}
    \frac{\partial z^\prime}{\partial x}=6z(1+\omega)~~~~\text{and}
\end{equation}
\begin{equation}
    \frac{\partial z^\prime}{\partial z}=6z+6(x\omega+x+z-1)~~~.
\end{equation}
At the critical point $(0,0)$, the Jacobian matrix becomes,

\begin{equation}
    J\Bigm\vert_{(0,0)}=\begin{bmatrix}
-3(1+\omega) & 0 \\
0 & -6 
\end{bmatrix}~~~.
\end{equation}
To obtain the eigenvalues of the above Jacobian matrix, we have 
\begin{equation}
    |J-\lambda I |\Bigm\vert_{(0,0)} = 0~~~
    \implies \lambda = -3(1+\omega) ~~\text{and}~~ -6 ~~~,~~~ \forall \omega~~~.
\end{equation}
Calculating the Jacobian matrix at the critical point $(1,0)$,

\begin{equation}
    J\Bigm\vert_{(1,0)}=\begin{bmatrix}
0 & -6\omega \\
0 & 6\omega 
\end{bmatrix}~~~.
\end{equation}
To obtain the eigenvalues of the above Jacobian matrix, we have 
\begin{equation}
    |J-\lambda I |\Bigm\vert_{(1,0)} = 0 ~~~
    \implies \lambda = 0 ~~\text{and} ~~ 6\omega ~~~,~~~ \forall \omega~~~.
\end{equation}

\subsection{Model-III : $f(Q)=\alpha Q^2+ v Q^2 \log (Q)$}

\begin{equation}
    x^\prime = 3x \left[\frac{2x(\omega+1)\{16x+13(z-1)\}}{(z-1)\{40x+33(z-1)\}} + 2(x\omega+x+z)-\omega-1\right] ~~~~~~~ \text{and}
\end{equation} 
\begin{equation}
    z^\prime = 6 z (x \omega +x+z-1)~~~.
\end{equation}

The critical points are $(0,0)$ , $\left(\frac{3}{4},0\right)$ and $\left(\frac{11}{12},0\right)$.
Now, solving the Jacobian matrix ($2$D System), 
\begin{equation}
\begin{split}
    \frac{\partial x^\prime}{\partial x} =3x\left[2(1+\omega)-\frac{80x\{16x+13(z-1)\}(1+\omega)}{\{40x+33(z-1)\}^2(z-1)} +\frac{32x(1+\omega)}{\{40x+33(z-1)\}(z-1)}+\frac{2\{16x+13(z-1)\}(1+\omega)}{\{40x+33(z-1)\}(z-1)}\right]\\
    +3\left[-1-\omega + \frac{2x\{16x+13(z-1)\}(1+\omega)}{\{40x+33(z-1)\}(z-1)}+2(x+z+x\omega)\right]~~,
\end{split}
\end{equation}
\begin{equation}
     \frac{\partial x^\prime}{\partial z} = 3x\left[2- \frac{2x\{16x+13(z-1)\}(1+\omega)}{\{40x+33(z-1)\}(z-1)^2}-\frac{66x\{16x+13(z-1)\}(1+\omega)}{\{40x+33(z-1)\}^2(z-1)}+\frac{26x(1+\omega)}{\{40x+33(z-1)\}(z-1)}\right]~~~,
\end{equation}
\begin{equation}
    \frac{\partial z^\prime}{\partial x}=6z(1+\omega)~~~\text{and}
\end{equation}
\begin{equation}
\frac{\partial z^\prime}{\partial z}=6z+6(x\omega+x+z-1)~~~.
\end{equation}
Calculating the Jacobian matrix at the critical point $(0,0)$,

\begin{equation}
    J\Bigm\vert_{(0,0)}=\begin{bmatrix}
-3(1+\omega) & 0 \\
0 & -6 
\end{bmatrix}~~~.
\end{equation}
To obtain the eigenvalues of the above Jacobian matrix, we have 
\begin{equation}
    |J-\lambda I |\Bigm\vert_{(0,0)} = 0 ~~~
    \implies \lambda = -3(1+\omega) ~~\text{and}~~ -6,~~~,~~~ \forall \omega~~~.
\end{equation}
Calculating the Jacobian matrix at the critical point $(\frac{3}{4},0)$,

\begin{equation}
    J\Bigm\vert_{\left(\frac{3}{4},0\right)}=\begin{bmatrix}
   6(1+\omega) & \frac{45}{8}(1+\omega) \\
   0 & \frac{3}{2}(3\omega-1)
    \end{bmatrix}~~~.
\end{equation}
The eigenvalues of the above Jacobian matrix are defined as,
\begin{equation}
    |J-\lambda I |\Bigm\vert_{\left(\frac{3}{4},0\right)} = 0 ~~~
    \implies \lambda = 6(1+\omega)~~\text{and} ~~\frac{3}{2}(3\omega-1)~~~,~~~ \forall \omega~~~.
\end{equation}
Calculating the Jacobian matrix at the critical point $\left(\frac{11}{12},0\right)$,

\begin{equation}
    J\Bigm\vert_{(\frac{11}{12},0)}=\begin{bmatrix}
   6(1+\omega) & \frac{11}{4}\{2+\frac{1}{6}(1+\omega)\} \\
   0 & \frac{1}{2}(11\omega -1)
    \end{bmatrix}~~~.
\end{equation}
Now, for obtaining the eigenvalues of the above Jacobian matrix, we have 
\begin{equation}
    |J-\lambda I |\Bigm\vert_{\left(\frac{11}{12},0\right)} = 0 ~~~
    \implies \lambda = 6(1+\omega)~~\text{and} ~~\frac{1}{2}(11\omega -1)~~~,~~~ \forall \omega~~~.
\end{equation}

\subsection{Model-IV : $f(Q)=\eta Q_0 \sqrt{\frac{Q}{Q_0}} \log\left(\frac{\lambda  Q_0}{Q}\right)$}
\begin{equation}
    x^\prime = 3 x \left\{\frac{x (\omega +1)}{3-4x-3z}+2 (x \omega +x+z)-\omega -1\right\}~~~~~\text{and}
\end{equation}
\begin{equation}
    z^\prime =6 z (x \omega +x+z-1)~~~.
\end{equation}

The critical points are $(0,0)$, $\left(\frac{3}{8},0\right)$ and $(1,0)$.
To solve the Jacobian matrix, we need to calculate the following :
\begin{equation}
    \frac{\partial x^\prime}{\partial x} =3x\left\{2(1+\omega)+\frac{4x(1+\omega)}{(3-4x-3z)^2}+\frac{1+\omega}{3-4x-3z}\right\}+3\left\{-1-\omega+\frac{x(1+\omega)}{3-4x-3z}+2(x+z+x\omega)\right\}~~~,
\end{equation}
\begin{equation}
    \frac{\partial x^\prime}{\partial z} =3x\left\{2+\frac{3x(1+\omega)}{(3-4x-3z)^2}\right\}~~~,
\end{equation}
\begin{equation}
    \frac{\partial z^\prime}{\partial x}=6z(1+\omega)~~~\text{and}
\end{equation}
\begin{equation}
    \frac{\partial z^\prime}{\partial z}=6z+6(x\omega+x+z-1)~~~.
\end{equation}
At the critical point $(0,0)$, the Jacobian matrix is 

\begin{equation}
    J\Bigm\vert_{(0,0)}=\begin{bmatrix}
-3(1+\omega) & 0 \\
0 & -6 
\end{bmatrix}~~~.
\end{equation}
Eigenvalues of the above Jacobian matrix are obtained by,
\begin{equation}
    |J-\lambda I |\Bigm\vert_{(0,0)} = 0 ~~~
    \implies \lambda = -3(1+\omega) ~~\text{and}~~ -6,~~~,~~~ \forall \omega~~~.
\end{equation}
Calculating the Jacobian matrix at the critical point $\left(\frac{3}{8},0\right)$,

\begin{equation}
    J\Bigm\vert_{\left(\frac{3}{8},0\right)}=\begin{bmatrix}
\frac{15}{4}(1+\omega) & \frac{9}{8}\left(2+\frac{1}{2}(1+\omega)\right) \\
0 & \frac{3}{4}(3\omega-5) 
\end{bmatrix}~~~.
\end{equation}
To obtain the eigenvalues of the above Jacobian matrix, we have 
\begin{equation}
    |J-\lambda I |\Bigm\vert_{\left(\frac{3}{8},0\right)} = 0 ~~~
    \implies \lambda = \frac{15}{4}(1+\omega) ~~\text{and}~~ \frac{3}{4}(3\omega-5) ~~~,~~~ \forall \omega~~~.
\end{equation}
Calculating the Jacobian matrix at the critical point $(1,0)$,

\begin{equation}
    J\Bigm\vert_{(1,0)}=\begin{bmatrix}
15(1+\omega) & 6+9(1+\omega) \\
0 & 6\omega 
\end{bmatrix}~~~.
\end{equation}
The eigenvalues of the above Jacobian matrix are defined as,
\begin{equation}
    |J-\lambda I |\Bigm\vert_{(1,0)} = 0 
    ~~~\implies \lambda = 6\omega ~~\text{and}~~ 15(1+\omega) ~~~,~~~ \forall \omega~~~.
    \end{equation}

\section{Appendix-II}

\subsection{Model I}
The non-metricity scalar is given by
\begin{equation}
    Q = 2 \sum H_i H_j = 6 H^2 - \sigma^2~~,
\end{equation}

The dimensionless density parameters are taken as
\begin{equation}
    x= \frac{\kappa \rho}{6 m n Q^{n-1} H^2}~,~~~~y = -\frac{m Q^{n}}{12 m n Q^{n-1} H^2}=\frac{1}{12 n Q H^2}~~~\text{and}~~~~z = \frac{\sigma^2}{6H^2}=\frac{2 n (x-1)+1}{1-2 n} = 1 - \frac{2n x }{2n-1}~~.
\end{equation}
The energy conservation equation is,
\begin{equation}
    x+y+z =1 .
\end{equation}

Using the energy conservation equation the expression of $x^\prime$ and $z^\prime$ reduces to
\begin{equation}
    x^\prime = 3 x \left\{{\frac{
   2 x (1 + \omega)}{(2 + \frac{1}{n-1}) (z-1)} + 
   2 (x + z + x \omega)}-1-\omega\right\}~~~ \text{and}
\end{equation}
\begin{equation}
    z^\prime = 6 z (x \omega +x+z-1)~~~.
\end{equation}

\subsection{Model II}
The non-metricity scalar is given by
\begin{equation}
    Q = 2 \sum H_i H_j = 6 H^2 - \sigma^2~~,
\end{equation}

The dimensionless density parameters are taken as
\begin{equation}
    x= \frac{\kappa \rho}{6 n e^{n Q} H^2}~,~~~~y = -\frac{e^{n Q}}{12 n e^{n Q} H^2}=\frac{1}{12 n H^2}~~~\text{and}~~~~z = \frac{\sigma^2}{6H^2} ~~.
\end{equation}
The energy conservation equation is,
\begin{equation}
    x+y+z =1 .
\end{equation}

Using the energy conservation equation the expression of $x^\prime$ and $z^\prime$ reduces to
\begin{equation}
    x^\prime = \frac{6 x \left\{x^2 (\omega +1)-2 x (\omega +1)+x (2 \omega +3) z+(z-1) (2 z-1-\omega)\right\}}{x+2 z-2}~~~\text{and}
\end{equation}
\begin{equation}
    z^\prime = 6 z (x \omega +x+z-1)~~~.
\end{equation}

\subsection{Model III}
The non-metricity scalar is given by
\begin{equation}
    Q = 2 \sum H_i H_j = 6 H^2 - \sigma^2~~,
\end{equation}

The dimensionless density parameters are taken as
\begin{equation}
    x= \frac{\kappa \rho}{6 \{(2 \alpha + v) Q+2 Q v \log (Q)\} H^2}~,~~~~y = -\frac{\alpha Q^2+ v Q^2 \log (Q)}{12 \{(2 \alpha + v) Q+2 Q v \log (Q)\} H^2}~~~\text{and}~~~~z = \frac{\sigma^2}{6H^2} ~~.
\end{equation}
The energy conservation equation is,
\begin{equation}
    x+y+z =1 .
\end{equation}

Using the energy conservation equation the expression of $x^\prime$ and $z^\prime$ reduces to
\begin{equation}
    x^\prime = 3x \left[\frac{2x(\omega+1)\{16x+13(z-1)\}}{(z-1)\{40x+33(z-1)\}} + 2(x\omega+x+z)-\omega-1\right] ~~~~~~~ \text{and}
\end{equation} 
\begin{equation}
    z^\prime = 6 z (x \omega +x+z-1)~~~.
\end{equation}

\subsection{Model IV}
The non-metricity scalar is given by
\begin{equation}
    Q = 2 \sum H_i H_j = 6 H^2 - \sigma^2~~,
\end{equation}

The dimensionless density parameters are taken as
\begin{equation}
    x= \frac{\kappa \rho}{6 \left[{\frac{\eta}{2} \sqrt{\frac{Q_0}{Q}} \left\lbrace{\log \left(\frac{\lambda  Q_0}{Q}\right)-2}\right\rbrace}\right] H^2}~,~~~~y = -\frac{\eta Q_0 \sqrt{\frac{Q}{Q_0}} \log\left(\frac{\lambda  Q_0}{Q}\right)}{12 \left[{\frac{\eta}{2} \sqrt{\frac{Q_0}{Q}} \left\lbrace{\log \left(\frac{\lambda  Q_0}{Q}\right)-2}\right\rbrace}\right] H^2}~~~\text{and}~~~~z = \frac{\sigma^2}{6H^2} ~~.
\end{equation}
The energy conservation equation is,
\begin{equation}
    x+y+z =1 .
\end{equation}

Using the energy conservation equation the expression of $x^\prime$ and $z^\prime$ reduces to
\begin{equation}
    x^\prime = 3 x \left\{\frac{x (\omega +1)}{3-4x-3z}+2 (x \omega +x+z)-\omega -1\right\}~~~~~\text{and}
\end{equation}
\begin{equation}
    z^\prime =6 z (x \omega +x+z-1)~~~.
\end{equation}

\section{Appendix-III}

\begin{table}[h!]
\centering
\caption{Qualitative summary of theoretical viability conditions for the four $f(Q)$ models.}

\begin{tabular}{|lcccc|}
\hline
Model & Ghost-free & $c_s^2 \ge 0$ & Late-time attractor & Viable region \\
\hline
Model I & Limited & Limited & Yes & Narrow ($n\simeq1$) \\
Model II & Moderate & Moderate & Yes & Restricted \\
Model III & Restricted & Restricted & Conditional & Small \\
Model IV & Yes & Yes & Yes & Broad \\
\hline
\end{tabular}
\end{table}

%%%%%%%%%%%%%%%%%%%%%%%%%%%%%%%%%%%%%%%%%%%%%%%%%%%%%%%%%%%%%%%%%%%%%%%%%%%%%%%%%%%
%--------------------------------------------------------------------
%	BIBLIOGRAPHY
%--------------------------------------------------------------------
\newpage

\bibliographystyle{ieeetr} % Title is link if provided

\bibliography{references1}

\end{document}